\documentclass[10pt,prd,aps,a4paper,notitlepage,nofootinbib]{revtex4-1}

\usepackage{graphicx}
\usepackage{mathrsfs}
\usepackage{amsmath,amsfonts,amssymb}
\usepackage{amsthm}
\usepackage{hyperref}
\usepackage{comment}
\theoremstyle{remark}

\makeatletter
\def\@bibdataout@rev{%
 \immediate\write\@bibdataout{%
  @CONTROL{%
   REVTEX42Control%
   \eprint@enable@sw{}{,eprint="1"}%
  }%
 }%
 \if@filesw
  \immediate\write\@auxout{\string\citation{REVTEX42Control}}%
 \fi
}%
\makeatother

\newcommand{\be}{\begin{equation}}
\newcommand{\ee}{\end{equation}}
\newcommand{\bea}{\begin{eqnarray}}
\newcommand{\eea}{\end{eqnarray}}
\newcommand{\bel}{\begin{align}}
\newcommand{\eel}{\end{align}}
\newcommand{\scri}{{\mathscr{I}^+}}

\def\p{\partial}

\def\i{{\rm i}}

\def\GMc2{{\rm G M_{\odot} c^{-2}}}

\def\vphi{\varphi}

\def\O{\mathcal{O}}

\begin{document}

\title{Semilinear wave equations in homothetic hyperboloidal coordinates and tail decay} 

\author{An{\i}l \surname{Zengino\u{g}lu}$^1$}
\email{anil@umd.edu}
\author{Sebastiano \surname{Bernuzzi}$^2$}
\email{sebastiano.bernuzzi@uni-jena.de}
\author{Andrea \surname{N{\"u}tzi}$^3$}
\email{andrea.nuetzi@math.su.se}
\affiliation{$^1$Institute for Physical Science and Technology, University of Maryland, College Park, MD 20742, USA}
\affiliation{$^2$Theoretisch-Physikalisches Institut, Friedrich-Schiller-Universit{\"a}t Jena, 07743, Jena, Germany}
\affiliation{$^3$Department of Mathematics, Stockholm University, Stockholm, Sweden}

\date{\today}

\begin{abstract}
Late-time wave tails decay at different rates along future null infinity and along timelike worldlines at finite radius. A compactified numerical evolution must represent both the slower decay at null infinity and the faster interior decay, producing an increasingly sharp transition between the two regimes. We address this difficulty for semilinear wave equations in Minkowski spacetime using homothetic hyperboloidal coordinates adapted to the scaling structure of the tail. In these coordinates, the tail approaches a smooth radial profile with the same decay rate at every compactified radius. The formulation therefore avoids the steepening of the radial profile seen in stationary hyperboloidal evolutions, and it reaches late times in a number of
steps that grows only logarithmically with retarded time. We demonstrate this approach using pseudospectral simulations in 3+1 dimensions and reproduce the generic decay rates conjectured by Rinne. We also provide numerical evidence consistent with a nongeneric codimension-one cancellation of the leading tail coefficient at null infinity, resulting in a faster decay rate.
\end{abstract}

\maketitle


\section{Introduction}
\label{sec:intro}

The late-time behavior in wave propagation problems is often dominated by power-law decay, referred
to
as tails.  They arise already at the linear level from backscattering
off the effective potential and the associated branch cut in the
frequency-domain Green function, as first described in Price's
classic analysis of perturbations of gravitational collapse
\cite{Price:1972pw}.  In nonlinear and semilinear wave equations,
tails can also be sourced by self-interaction. The nonlinear terms act as an
effective long-time source that produces universal polynomial decay
\cite{Bizon:2008ew, Szpak:2008jv}.  Because tails encode long-range
dynamics and asymptotic structure, they provide a stringent test case
for comparing numerical evolution with analytical predictions
\cite{Harms:2013ib}.
There has recently been renewed attention on tails driven by the
rapidly improving accuracy of gravitational waveform modeling
\cite{Albanesi:2023bgi,DeAmicis:2024not,Islam:2024vro,
Islam:2025wci,Alnasheet:2025mtr,Vega:2026lfs}.
For example, \cite{DeAmicis:2024eoy} explored late-time
gravitational-wave tails directly in fully nonlinear $3+1$
numerical-relativity simulations of merging black holes and argued
that the tail signal can be substantially more prominent than
previously anticipated. Semilinear tails at finite radii have also been investigated in black-hole
spacetimes 
\cite{Ling:2025wfv,Ling:2026ynd}.

A challenging aspect of computing accurate tail decay rates
numerically is that they are asymptotic in time.  Their computation
requires highly accurate evolution schemes over long times, so
standard truncation methods are computationally limited in this
problem.  Hyperboloidal methods address this by evolving on spacelike
hypersurfaces that asymptote to $\scri$, allowing one to include null
infinity in the computational domain and avoiding artificial timelike
outer boundaries \cite{Zenginoglu:2008wc}.  In a scri-fixing gauge,
the coordinate location of $\scri$ is time-independent, so $\scri$ is
represented by the same grid points throughout the evolution and
radiation can be extracted on a known grid boundary
\cite{Zenginoglu:2007jw}.  Hyperboloidal evolution therefore provides
a clean and robust framework for computing tails and other late-time
asymptotics \cite{Zenginoglu:2008uc,
Zenginoglu:2009ey, Jasiulek:2011ce, Racz:2011qu, Bernuzzi:2012ku,
Zenginoglu:2012us, Harms:2013ib, Harms:2014dqa,
Csukas:2019kcb}.
Full $3+1$ time-domain implementations on Minkowski spacetime include
scalar-wave evolutions with hyperboloidal spectral methods, nonlinear
source terms, and no imposed spatial symmetries
\cite{Zenginoglu:2010cq,Zenginoglu:2010zm,Hilditch:2016xzh, Gautam:2021ilg, Peterson:2023bha, Rinne:2025, Reddy:2026dfz}.

A complication in hyperboloidal studies of tail decay rates is
that tail exponents depend on the approach to timelike infinity.  The
decay rate measured along $\scri$ differs from the rate measured along
timelike worldlines at finite radii.  For example, for a scalar
multipole of angular momentum $\ell$ in Schwarzschild spacetime, one
expects a faster decay at finite radius than at null infinity, and the
difference grows with $\ell$ \cite{Gundlach:1993tp}.  From a numerical
perspective, this observer dependence leads to a \emph{long-time
resolution problem} on the compactified grid.  When the field decays
more slowly near $\scri$ than in the interior, the late-time solution
develops an increasingly sharp radial transition between regions
governed by different exponents; see, for example, Fig.~5 in
\cite{Zenginoglu:2008wc}.  The same mechanism appears for semilinear
wave equations \cite{Rinne:2025}.  For sufficiently long runtimes, one
must then use non-uniform grids, adaptive mesh refinement, or another
resolution strategy to maintain accuracy.

This paper presents an exponentially efficient computational method
adapted to the asymptotic tail regime for
semilinear wave equations.  We do this by representing the conformal
Minkowski metric in \emph{homothetic coordinates} adapted to the
scaling symmetry (self-similarity) of the semilinear equation. Such self-similar coordinates have
played a central role in the analysis of semilinear wave equations in the mathematical literature
\cite{Donninger:2013sba, Burtscher:2015, Bonk:2026}. A
key difference to stationary hyperboloidal compactifications is that
the conformal factor is time-dependent and the relation to retarded Killing time at $\scri$ is
exponential, so homothetic time covers exponentially large intervals of retarded time.

This time compression is possible because the tail remains resolved in
homothetic coordinates.  A self-similar tail separates into an
exponentially decaying factor \(e^{-\tilde q_\ell\tau}\) and a smooth
radial profile, with the same exponent at every fixed compactified
radius, $\rho\in(0,1]$. Therefore, the
homothetic formulation both reaches
late retarded times with logarithmic cost and avoids the
increasingly steep radial transition that limits stationary
compactifications.  We validate the method numerically using 3+1 pseudospectral evolutions and
compare the
measured decay rates against analytical predictions.

The paper has three related contributions.  First, it provides a
numerical 3+1 implementation of a homothetic hyperboloidal formulation of the conformal semilinear
wave
equation, including the corresponding balance laws.  Second, it
demonstrates numerically that tail extraction in this formulation has
logarithmic scaling in the asymptotic regime: homothetic
time reaches retarded time $u=e^\tau$, so uniform $\tau$-steps cover
exponentially large intervals in $u$, while the self-similar tail
profile remains smooth on the compactified grid.  Third, it  tests the generic character of the
conjectured tail decay rates. Specifically, we construct a one-parameter family of
axisymmetric data for the cubic wave equation and bisect in that family to cancel the leading
coefficient at $\scri$.  The tuned monopole data decay at $\scri$ with
the subleading exponent $q=2$, whereas untuned monopole data recover
the expected generic exponent $q=1$.

The paper is organized as follows.  In Sec.~\ref{sec:wave} we review
the semilinear wave equation and its conformal transformation.  In
Sec.~\ref{sec:stationary} we describe stationary hyperboloidal
coordinates and their energy balance, while Sec.~\ref{sec:homothetic}
introduces homothetic hyperboloidal coordinates.  In
Sec.~\ref{sec:rates} we explain why the homothetic formulation gives
the same decay exponent at every compactified
radius.  In
Sec.~\ref{sec:results} we validate the implementation, compare
stationary and homothetic tail extraction and computational cost,
study non-compact data at $\scri$ and compact-data higher multipoles,
and construct the tuned cubic monopole example.  We conclude in
Sec.~\ref{sec:conclusions}.  Appendix~\ref{app:numerics} describes the
spectral implementation and the treatment of regularity at the
origin.


\section{Semilinear wave equation and conformal transformation}
\label{sec:wave}

\subsection{The wave equation}

The Minkowski metric in standard spherical coordinates is
\be
\eta = -dt^2 + dr^2 + r^2 d\sigma^2,
\ee
where $\sigma$ denotes angular coordinates on $\mathbb S^2$ and $d\sigma^2$ is the standard
unit-sphere line element.  We denote spatial Cartesian
coordinates by $x^i$ with
$r=|x|=\sqrt{\delta_{ij}x^i x^j}$.  Writing
$x^\mu=(t,x^i)$, we use the metric signature $(-,+,+,+)$.  Greek
indices $\mu,\nu,\ldots$ run over spacetime components $0,\ldots,3$,
lowercase Latin indices $i,j,\ldots$ run over spatial components
$1,2,3$, and uppercase Latin indices $A,B,\ldots$ denote angular
components on $\mathbb S^2$.
We consider the focusing semilinear wave equation in Minkowski space,
\be
\Box_\eta \phi(t,x) = \left(-\partial_t^2 + \Delta\right)\phi(t,x)
= -\,|\phi(t,x)|^{p-1}\phi(t,x),
\ee
where the minus sign corresponds to the focusing nonlinearity.  We
consider only odd integer $p$ and real-valued $\phi$, so the
right-hand side can be written as $-\phi^p$ and the equation becomes
\be\label{eq:standard_wave}
\Box_\eta \phi + \phi^p=0, \qquad p\in\{3,5,7,\dots\}.
\ee
The most extensively studied case
is the cubic equation, $p=3$, because of its conformal invariance.
Hyperboloidal methods were applied to study the cubic wave equation
both numerically \cite{Bizon:2008zd} and analytically
\cite{Donninger:2013sba}.  It was also used as a numerical testbed for
hyperboloidal implementations in $3+1$ dimensions
\cite{Zenginoglu:2010cq, Zenginoglu:2010zm}.  Going beyond the cubic
case, \citet{Rinne:2025} recently provided an extended numerical study
of the semilinear wave equation on hyperboloidal slices for different powers and modes.

The semilinear wave equation has stress-energy tensor given by
\be
T_{\mu\nu}
=
\p_\mu\phi\,\p_\nu\phi
-\frac12\eta_{\mu\nu}\eta^{\alpha\beta}\p_\alpha\phi\,\p_\beta\phi
+\frac{1}{p+1}\,\eta_{\mu\nu}\,\phi^{p+1}.
\ee
The stress-energy tensor yields the conserved current
$J^\mu=T^\mu{}_{\nu}(\p_t)^\nu$ satisfying $\p_\mu J^\mu=0$.
The corresponding energy is
\be\label{eq:E}
E_p(t)
=
\int d^3 x \left(
\frac12 \left(\p_t \phi \right)^2
+\frac12 |\nabla\phi|^2
-\frac{1}{p+1}|\phi|^{p+1}
\right),
\qquad
\frac{d}{dt}E_p(t)=0,
\ee
so that $E_p(t)=E_p(0)$ for regular solutions.
Because the energy is not positive definite, the focusing semilinear
wave equation allows for blow-up solutions.  Critical phenomena then
appear at the threshold between blow-up and decay.  The global energy \eqref{eq:E}
is not only conserved in the sense of satisfying a balance law; it is
constant in time.

\subsection{The conformal transformation}
To include null infinity in the numerical domain, we perform a
Penrose-style conformal compactification \cite{Penrose:1962ij}.  It is natural in this
setting to consider the conformal wave equation. Let $n$ be the spacetime dimension and let
$g=\Omega^{2}\eta$ be a conformal rescaling of Minkowski space
$(\mathbb R^{1,n-1},\eta)$.  In this subsection, Greek indices run
over $0,\ldots,n-1$.  The conformal
factor $\Omega$ is smooth with $\Omega>0$ on the physical region and
$\Omega = 0, \ d\Omega\neq 0$ at $\scri$.  The Lorentzian Yamabe
operator reads
\begin{equation}
  P_g := \Box_g - \frac{n-2}{4(n-1)}\,\mathcal R[g],
  \label{eq:yamabe-operator-n}
\end{equation}
where $\mathcal{R}$ denotes the scalar curvature. The wave operator transforms as
\begin{equation}
  \Box_{\eta}\,\phi =
  \Omega^{\frac{n+2}{2}}\,
  P_g\left(\Omega^{-\frac{n-2}{2}}\,\phi\right).
  \label{eq:conformal-covariance-n}
\end{equation}
Introducing the conformal field with conformal weight $\frac{n-2}{2}$
\begin{equation}
  \psi := \Omega^{-\frac{n-2}{2}}\,\phi,
  \label{eq:conformal-field-n}
\end{equation}
the physical semilinear wave equation \eqref{eq:standard_wave}
is equivalent to the conformal semilinear equation
\begin{equation}
  \left(\Box_g-\frac{n-2}{4(n-1)}\mathcal R[g]\right)\psi
   + \,\Omega^{\frac{(n-2)p-(n+2)}{2}}\,\psi^p = 0.
  \label{eq:conformal-semilinear-n}
\end{equation}
The scalar curvature $\mathcal R[g]$ is given by
\begin{equation}
  \mathcal R[g] =
  - 2(n-1)\,\Omega^{-3}\,\Box_{\eta}\Omega
  -
  (n-1)(n-4)\,\Omega^{-4}\,\eta^{\mu\nu}\partial_{\mu}\Omega\,\partial_{\nu}\Omega.
  \label{eq:Ricci-conformal-n-Omega}
\end{equation}
In four spacetime dimensions ($n=4$), only the first term survives and
$\mathcal R[g] =-6 \Omega^{-3} \Box_{\eta}\Omega$.  The conformal
weight becomes $\frac{n-2}{2}=1$, so the conformal rescaling is
simply $\psi := \Omega^{-1}\phi$.
For the scri-fixing compactifications used below, approaching $\scri$
at fixed retarded time gives $\Omega\sim r^{-1}$, up to a smooth
nonvanishing factor.  Therefore, the conformal field
$\psi=\Omega^{-1}\phi$ agrees asymptotically at $\scri$, up to such a
factor, with the radiation field $r\phi$.  The conformal semilinear
equation reads
\begin{equation}\label{eq:conformal-semilinear}
  \Box_g \psi +\Omega^{-3}\,(\Box_{\eta}\Omega)\, \psi
  + \Omega^{p-3} \psi^p = 0.
\end{equation}
We formulate and solve this equation numerically in two
hyperboloidal scri-fixing coordinate systems: the stationary
coordinates described in Sec.~\ref{sec:stationary} and the homothetic
coordinates described in Sec.~\ref{sec:homothetic}.


\section{Stationary hyperboloidal coordinates}\label{sec:stationary}
In this section, we review the construction of stationary
hyperboloidal coordinates and the treatment of the semilinear wave
equation in these coordinates.  We derive the energy
balance law for the associated Noether quantity.

\subsection{Scri-fixing through time-translation invariance}

Penrose's conformal compactification translates asymptotic analysis
into local analysis at the conformal boundary \cite{Penrose:1962ij}.  This
geometric framework also enables the numerical solution of wave
equations on unbounded domains \cite{Frauendiener:2000mk}.  However, future null infinity is
an ingoing null hypersurface in Penrose-type coordinates, which can
lead to loss of numerical accuracy or complications related to the
boundary treatment beyond null infinity. 
A convenient choice for numerical computations is scri-fixing, in
which $\scri$ is mapped to a fixed coordinate location and becomes a
purely outflow surface for the interior evolution problem.  The first
hyperboloidal scri-fixing coordinates in Minkowski spacetime were
constructed by Gowdy \cite{Gowdy:1981}.  In this construction,
time-translated hyperboloids of constant negative curvature foliate
spacetime.  The approach was largely overlooked at the time, but the hyperboloidal
approach in Minkowski spacetime was studied independently in later works \cite{Moncrief:2000,
Fodor:2003yg} motivated by the hyperboloidal initial value problem for Einstein equations
\cite{Friedrich:1983,Frauendiener:2000mk}.  A general
construction of stationary hyperboloidal compactifications for
stationary background spacetimes was introduced in
\cite{Zenginoglu:2007jw}, which became the standard approach to
incorporating hyperboloidal coordinates in perturbative studies.

The construction of stationary scri-fixing in
\cite{Zenginoglu:2007jw} combines a hyperboloidal time
transformation, a spatial
compactification, and a conformal rescaling.  These choices satisfy
asymptotic conditions that ensure regularity of the metric across the
conformal boundary and respect the timelike Killing symmetry of the
background.  In particular, the height function, the compactification
map, and the conformal factor are chosen to be independent of the
Killing time. We define new coordinates $\tau$ and $\rho$ via
\begin{equation}\label{eq:tau_rho_def}
  \tau := t - h(r),
  \qquad
  \rho := g(r).
\end{equation}
Here $h$ and $g$ are monotone increasing functions where $g(0)=0$ and $g(r) \to \rho_{\mathscr I^+}$ as $r\to \infty$. Thus $\rho = 0$ represents the physical origin, while $\rho=\rho_{\scri}$ represents future null infinity. The concrete stationary compactification used in the numerical experiments is given in Sec.~\ref{sec:stationary-hyperboloids}. We further assume that the derivatives
\begin{equation}\label{eq:H_G_def}
  H(r) := \frac{dh}{dr},
  \qquad
  G(r) := \frac{dg}{dr},
\end{equation}
satisfy $H\to 1$ and $G \to 0$ as $r \to \infty$. A convenient choice that connects the compactification with the conformal rescaling is given by
\be\label{eq:compactification_choice}
r=\frac{\rho}{\Omega(\rho)} \ \implies \ G=\frac{\Omega^2}{L} \ \text{with} \  L:=\Omega-\rho
\Omega'.
\ee
Then the conformal Minkowski metric becomes in $(\tau,\rho,\theta,\varphi)$ coordinates:
\begin{equation}\label{eq:g_tau_rho_general}
  g = \Omega^2 \eta 
  = - G L\,d\tau^{2}
    -2\, H L\,d\tau\,d\rho
    +\frac{1-H^{2}}{G} L \,d\rho^{2}
    +\rho^{2}\,d\sigma^{2}.
\end{equation}
Let $\rho_{\scri}$ denote the coordinate location of future null
infinity. The conformal extension through $\rho=\rho_{\scri}$ is
regular provided the coefficient $\frac{1-H^{2}}{G}$ extends smoothly
to $\rho=\rho_{\scri}$. This requirement is the asymptotic condition
for scri-fixing hyperboloidal compactification in this setting. The
radial characteristic speeds read
\begin{equation}\label{eq:stationary_radial_speeds}
c_+ = \frac{G}{1-H}, \qquad c_-=-\frac{G}{1+H}.
\end{equation}
The asymptotic condition for scri-fixing implies that $c_+(\rho_{\scri})$ is non-vanishing finite
and $c_-(\rho_{\scri})$ vanishes, so $\scri$ is an outflow boundary for the interior evolution
problem.  The timelike Killing vector $\partial_t$ becomes $\partial_\tau$ in the new coordinates,
so the metric is manifestly time-independent.


\subsection{Conformal wave equation in stationary hyperboloidal coordinates}

We write the conformal wave equation \eqref{eq:conformal-semilinear} in the stationary hyperboloidal
compactification described above. The $(\tau,\rho)$-block determinant for the metric
\eqref{eq:g_tau_rho_general} is
$\det(g_{\mu\nu})_{\mu,\nu\in\{\tau,\rho\}}=-L^2$. The inverse metric
components read
\begin{equation}\label{eq:inverse_spherical}
  g^{\tau\tau}=-\frac{1-H^2}{GL},\qquad
  g^{\tau\rho}=g^{\rho\tau}=-\frac{H}{L},\qquad
  g^{\rho\rho}=\frac{G}{L},\qquad
  g^{AB}=\rho^{-2}\gamma^{AB}.
\end{equation}
We also have $\sqrt{-g}= \rho^2\sqrt{\gamma}\,L$.  Using that $L,H,\Omega$ depend on $\rho$
only, we get
\begin{equation}\label{eq:box_spherical}
  \Box_g \psi =
  \frac{1}{\sqrt{-g}}\partial_\mu\left(\sqrt{-g}\,g^{\mu\nu}\partial_\nu \psi \right) =
  -\frac{1-H^2}{GL}\,\psi_{\tau\tau}
  -\frac{2H}{L}\,\psi_{\tau\rho}
  +\frac{G}{L}\,\psi_{\rho\rho}
  - \frac{1}{L\,\rho^2} (\rho^2 H)'\,\psi_\tau
  + \frac{1}{L\,\rho^2} (\rho^2 G)' \psi_\rho
  +\frac{1}{\rho^2}\Delta_{\mathbb S^2}\psi,
\end{equation}
where primes denote derivatives with respect to $\rho$, and
subscripts on $\psi$ denote partial derivatives.
With the compactification choice \eqref{eq:compactification_choice}, we have for the Ricci term 
\be\label{eq:ricci-term} 
\Omega^{-3} \Box_\eta \Omega = \Omega^{-3} \frac{1}{r^2} \partial_r \left(r^2 \partial_r \Omega\right) = \frac{\Omega}{\rho^2 L} \partial_\rho \left(\frac{\rho^2}{L} \partial_\rho \Omega \right).
\ee
The conformal semilinear wave equation in stationary hyperboloidal coordinates becomes
\begin{equation}\label{eq:conformal_stationary}
\begin{aligned}
0 & = \Box_g \psi +\Omega^{-3}\,(\Box_{\eta}\Omega)\, \psi
  + \Omega^{p-3} \psi^p \\
& =  -\frac{1-H^2}{GL}\,\psi_{\tau\tau}
  -\frac{2H}{L}\,\psi_{\tau\rho}
  +\frac{G}{L}\,\psi_{\rho\rho}
  - \frac{1}{L\,\rho^2} (\rho^2 H)'\,\psi_\tau
  + \frac{1}{L\,\rho^2} (\rho^2 G)' \psi_\rho 
  +\frac{1}{\rho^2}\Delta_{\mathbb S^2}\psi\\
  & \qquad + \frac{\Omega}{\rho^2 L} \partial_\rho \left(\frac{\rho^2}{L} \partial_\rho \Omega \right) \psi  + \Omega^{p-3} \psi^p.
\end{aligned}
\end{equation}

\subsection{Energy balance}
The conformal Noether balance can be derived either from the
stress-energy tensor of the conformal wave equation or directly from
the equation \eqref{eq:conformal_stationary}. Multiplying
\eqref{eq:conformal_stationary} by
$L(\rho)\rho^2\,\psi_\tau$ and integrating by parts in $\tau$, $\rho$,
and on $\mathbb S^2$ yields the local conservation law
\begin{equation}\label{eq:local-conservation}
  \partial_\tau e + \partial_\rho j_\rho + \nabla_{\mathbb S^2}\cdot j_{\mathbb S^2} = 0,
\end{equation}
where $j_\rho,j_{\mathbb S^2}$ indicate the radial and angular components of the
flux vector.
Let $|\nabla_{\mathbb S^2}\psi|^2:=\gamma^{AB}\,(\nabla_A\psi)(\nabla_B\psi)$, where $\gamma_{AB}$
is the round metric on $\mathbb S^2$. The corresponding densities are
\begin{align}\label{eq:energy-densities}
  e(\tau,\rho,\sigma)
  &=
  \frac{\rho^2}{2}\left(
    \frac{1-H^2}{G}\,\psi_\tau^2
    +
    G\,\psi_\rho^2
  \right)
  +
  \frac{L}{2}\,|\nabla_{\mathbb S^2}\psi|^2
  -
  L\rho^2\,\left(\frac12 \Omega^{-3}(\Box_\eta\Omega)\, \psi^2 + \frac{\Omega^{p-3}}{p+1}\psi^{p+1} \right),\\[0.5em]
  j_\rho(\tau,\rho,\sigma) 
  &=
  \rho^2\left(
    H\,\psi_\tau^2
    -
    G\,\psi_\tau\,\psi_\rho
  \right)
  =
  \rho^2\,\psi_\tau\left(H\psi_\tau - G\psi_\rho\right), \\[0.5em]
  j_{\mathbb S^2}(\tau,\rho,\sigma)
  &=
  -\,L\,\psi_\tau\,\nabla_{\mathbb S^2}\psi.
\end{align}
Integrating the local conservation law over $[0,\rho_{\scri}]\times\mathbb S^2$ and using the
definition of the conformal energy functional
\[
E(\tau):=\int_{0}^{\rho_{\scri}}\int_{\mathbb S^2} e(\tau,\rho,\sigma)\,d\sigma\,d\rho,
\]
where $d\sigma$ is the area form on $\mathbb S^2$,
we obtain
\begin{align}
\frac{d}{d\tau}E(\tau)
&=\int_{0}^{\rho_{\scri}}\int_{\mathbb S^2}\partial_\tau e\,d\sigma\,d\rho
= -\int_{0}^{\rho_{\scri}}\int_{\mathbb S^2}\left(\partial_\rho j_\rho+\nabla_{\mathbb S^2}\cdot
j_{\mathbb S^2}\right)\,d\sigma\,d\rho
\nonumber\\
&= -\int_{\mathbb S^2} j_\rho(\tau,\rho_{\scri},\sigma)\,d\sigma
+\int_{\mathbb S^2} j_\rho(\tau,0,\sigma)\,d\sigma
-\int_{0}^{\rho_{\scri}}\int_{\mathbb S^2}\nabla_{\mathbb S^2}\cdot j_{\mathbb S^2}\,d\sigma\,d\rho.
\label{eq:energy-balance-general}
\end{align}
The angular divergence integrates to zero for each fixed $\rho$, since $\mathbb S^2$ has no
boundary. The origin term drops off by regularity,
\[
j_\rho(\tau,0,\sigma)=\lim_{\rho\to0}\rho^2\left(H\psi_\tau^2-G\psi_\tau\psi_\rho\right)=0.
\]
At future null infinity $\rho=\rho_{\scri}$ we have $H\to 1$ and $G\to 0$. Hence the radiated flux
through $\scri$ is
\begin{equation}\label{eq:flux-at-scri}
j_\rho(\tau,\rho_{\scri},\sigma)
=
\rho_{\scri}^2\,\psi_\tau(\tau,\rho_{\scri},\sigma)^2,
\end{equation}
and therefore the corresponding balance quantity is monotone
non-increasing:
\begin{equation}\label{eq:stationary-energy-decay-updated}
\frac{d}{d\tau}E(\tau) =
-\rho_{\scri}^2\int_{\mathbb S^2}\psi_\tau(\tau,\rho_{\scri},\sigma)^2\,d\sigma
\le0.
\end{equation}
So for any $\tau_2>\tau_1$,
\begin{equation}\label{eq:integrated-energy-decay-updated}
E(\tau_2) = E(\tau_1) -
\rho_{\scri}^2\int_{\tau_1}^{\tau_2}\int_{\mathbb S^2}\psi_\tau(s,\rho_{\scri},\sigma)^2\,d\sigma\,ds.
\end{equation}
The explicit balance law is an essential feature and advantage of
compactification at null infinity and allows us to accurately measure
outgoing radiation while maintaining high accuracy for the remaining
bulk solution.

\subsection{Time-translated hyperboloids}
\label{sec:stationary-hyperboloids}
In recent years, many examples of stationary hyperboloidal
compactifications have been used in the literature; see
\cite{PanossoMacedo:2023qzp, Vano-Vinuales:2024tat,
Zenginoglu:2025sft} for reviews. Here, we consider
time-translated hyperboloids as in \cite{Gowdy:1981, Fodor:2003yg, Bizon:2008zd, Zenginoglu:2010zm,
Rinne:2025}. 
Such time-translated hyperboloids are defined by
$(\tau-t)^2-r^2=R^2$, where $R$ is the curvature radius of the
hyperboloids.  The first hyperboloidal scri-fixing compactification
introduced by Gowdy \cite{Gowdy:1981} relied on a stereographic
projection.  Here we follow the mapping of the hyperboloids to the
Poincar\'e ball described in \cite{Bizon:2008zd}.  Since we are
interested in a future hyperboloidal foliation, we choose the positive
root for the height function, $h=\sqrt{R^2+r^2}$.  For the Poincar\'e
ball mapping, we use the spatial compactification with conformal
factor $\Omega(\rho)=\frac{1-\rho^{2}}{2R}$:
\begin{equation}\label{eq:r_of_rho}
  r(\rho)= \frac{\rho}{\Omega} = \frac{2R\rho}{1-\rho^{2}},
  \qquad
  r'(\rho)=\frac{dr}{d\rho}= \frac{2R(1+\rho^{2})}{(1-\rho^{2})^{2}},
  \qquad
  G(\rho)=\frac{d\rho}{dr}=\frac{1}{r'(\rho)}=\frac{(1-\rho^{2})^{2}}{2R(1+\rho^{2})},
  \quad
  L(\rho)=\frac{1+\rho^2}{2R}.
\end{equation}
together with the hyperboloidal height function
\begin{equation}\label{eq:h_of_r}
  h(r)=\sqrt{R^{2}+r^{2}}=R\,\frac{1+\rho^{2}}{1-\rho^{2}},
  \qquad
  H(\rho)=\frac{dh}{dr}=\frac{r}{\sqrt{R^{2}+r^{2}}}=\frac{2\rho}{1+\rho^{2}}.
\end{equation}
The conformal metric \eqref{eq:g_tau_rho_general} takes the simple form
\begin{equation}\label{eq:conformal_metric_final}
  g
  = -\Omega^{2}\,d\tau^{2}
    -\frac{2\rho}{R}\,d\tau\,d\rho
    + d\rho^{2}
    + \rho^{2}\,d\sigma^{2}.
\end{equation}
All coefficients in \eqref{eq:conformal_metric_final} are smooth on $0\le \rho\le 1$.  At $\rho=1$
one has $\Omega=0$, but $g$ remains non-degenerate and smooth.

For the metric \eqref{eq:conformal_metric_final}, the Ricci term \eqref{eq:ricci-term} reads
\begin{equation}\label{eq:Ricci_scalar}
  \Omega^{-3} \Box_\eta \Omega = -\frac{2(1-\rho^{2})(3+\rho^{2})}{(1+\rho^{2})^{3}}.
\end{equation}
Inserting these expressions into \eqref{eq:conformal_stationary}, and multiplying by
$L^2=\frac{(1+\rho^2)^2}{4R^2}$, the conformal semilinear equation reads
\begin{equation}\label{eq:poincare_equation}
\begin{aligned}
0 =&
-\psi_{\tau\tau}
-\frac{2\rho}{R}\,\psi_{\tau\rho}
+\Omega^2\psi_{\rho\rho}
+\frac{(1+\rho^2)^2}{4R^2\,\rho^2}\,\Delta_{\mathbb S^2}\psi
-\frac{3+\rho^2}{R(1+\rho^2)}\,\psi_\tau
-\frac{\Omega(2\rho^4+3\rho^2-1)}{R\,\rho\,(1+\rho^2)}\,\psi_\rho\\[1mm]
&-\frac{\Omega(3+\rho^2)}{R(1+\rho^2)}\,\psi
+\frac{(1+\rho^2)^2}{4R^2}\Omega^{p-3}\psi^p \, .
\end{aligned}
\end{equation}

The in- and outgoing radial characteristics can be obtained by transformation from the Minkowski
null coordinates. We get
\begin{equation}\label{eqn:hyp_full_chars}
  u = t-r=\tau+R\,\frac{1-\rho}{1+\rho},
  \qquad
  v = t+r=\tau+R\,\frac{1+\rho}{1-\rho}.
\end{equation}
The characteristic speeds are
\begin{equation}\label{eq:stationary_chars}
c_+ = \frac{(1+\rho)^2}{2R}, \qquad c_-=-\frac{(1-\rho)^2}{2R}.
\end{equation}
These relations imply that the outer boundary is outflow: the outgoing
radial characteristic smoothly crosses the outer boundary with speed
$c_+|_{\scri}=2/R$, and no incoming characteristic enters from it as $c_-|_{\scri}=0$.

\section{Homothetic hyperboloidal coordinates}\label{sec:homothetic}
The stationary hyperboloidal compactification discussed in the
previous section is time-translation invariant.  An alternative
approach to hyperboloidal compactification is based on conformal
Killing vector fields.  In Minkowski spacetime, the dilatation operator
$D=x^\mu\partial_\mu=t\partial_t+r\partial_r$ generates scalings.
We can construct
a hyperboloidal foliation associated with dilatations that is not
time-translation invariant but is still scri-fixing
\cite{Donninger:2013sba,Burtscher:2015,Nutzi:2025kqc,Bonk:2026}.

\subsection{Scri-fixing through self-similarity}

The dilatation operator $D$ leaves the light cone invariant.
We construct a compactification of the future
timelike cone
$\mathcal{C}^+ := \{(t,\vec x)\in\mathbb{R}^{1+3}:\ t> |\vec x|\}$
based on the dilatation flow.  Introduce compactified spatial coordinates
\begin{equation}\label{eq:homothetic_compact_spatial}
  \vec\chi := \frac{\vec x}{t},\qquad \rho:=|\vec\chi|=\frac{r}{t}\in[0,1),
\end{equation}
which are invariants of the flow, $D(\vec\chi)=0$ and $D(\rho)=0$.
To parametrize the dilatation orbits we introduce a future-directed
hyperboloidal time coordinate
\begin{equation}\label{eq:homothetic_time_T_def}
  \tau := \ln\left(\frac{t^2-r^2}{2Rt}\right)
       = \ln\left(\frac{t(1-\rho^2)}{2R}\right),
\end{equation}
where $R>0$ is an arbitrary constant with dimensions of length.  The
factor $2R$ reflects a rescaling freedom along the dilatation flow:
replacing $\tau\mapsto \tau+c$ is equivalent to
$R\mapsto Re^{-c}$.  More generally, since $\rho$ and the angles are
invariants of $D$, we can reparametrize
$\tilde\tau=\tau+h(\rho,\sigma)$ without changing the homothetic
property $D=\partial_{\tilde\tau}$.  Such a transformation changes the
foliation of $\mathcal{C}^+$ but preserves the adaptation to
dilatations.  We use the length scale $R$ to control the relationship
between homothetic time and Bondi time.

It is instructive to compare homothetic coordinates with Milne
coordinates $(\tau_M,\xi,\sigma)$ on $\mathcal{C}^+$, with
$t=\tau_M\cosh\xi$, $r=\tau_M\sinh\xi$, and
$\tau_M=\sqrt{t^2-r^2}$.  The spatial coordinate is the same invariant,
$\rho=r/t=\tanh\xi$, and the dilation operator becomes
$D=\tau_M\partial_{\tau_M}=\partial_{\eta}$.  In these variables the
homothetic time differs from logarithmic Milne time by a
$\rho$-dependent shift,
$\tau= \ln\left(\frac{\tau_M}{2R}\right)
+ \frac12\ln(1-\rho^2)$.

The inverse transformation corresponding to
\eqref{eq:homothetic_compact_spatial} and
\eqref{eq:homothetic_time_T_def} is
\begin{equation}\label{eq:homothetic_inverse_map}
  t=\frac{2R\,e^{\tau}}{1-\rho^2},\qquad
  \vec x=\frac{2R\,e^{\tau}}{1-\rho^2}\,\vec\chi,\qquad
  r=\frac{2R\,\rho\,e^{\tau}}{1-\rho^2}.
\end{equation}
Thus the unbounded cone $\mathcal{C}^+$ is mapped to the cylinder
$\tau\in\mathbb{R}$ with spatial slices given by the open unit ball
$|\vec\chi|<1$.  Direct computation from
\eqref{eq:homothetic_inverse_map} yields
\[
  \eta = -dt^2 + d\vec x^{\,2}
  = t^2\left(-(1-\rho^2) d\tau^2 - 2 \rho\, d\rho\, d\tau + d\rho^2 + \rho^2 d\sigma^2\right).
\]
We choose the conformal factor as
\[
\Omega := \frac{1}{t}=\frac{1-\rho^2}{2R}\,e^{-\tau}.
\]
The conformal metric $g:=\Omega^2\eta$ extends smoothly to
$\rho=1$ and $\partial_\tau$ is Killing for $g$.  The conformal metric
describes the static patch of de Sitter spacetime in horizon-fixing
coordinates \cite{Parikh:2002qh}.  Future null infinity is mapped to
the fixed, time-independent boundary
$\scri \equiv \{\Omega=0\} \equiv \{\rho=1\}$.

The in and outgoing characteristics are
\begin{equation}\label{eq:homothetic_chars}
u = t - r = \frac{2R e^\tau}{1+\rho}, \qquad v = t+ r=\frac{2R e^\tau}{1-\rho}.
\end{equation}
In particular, at $\scri$, the relation between retarded Bondi time $u$ and homothetic time $\tau$ is given
by $u|_{\scri}=R e^\tau$. The characteristic speeds are independent of the overall scale $R$:
\[ c_+ = 1+\rho, \qquad c_- = -(1-\rho). \]
As in stationary hyperboloidal compactification, no incoming characteristic enters from $\rho=1$, so
no outer boundary condition is required.

\subsection{The conformal wave equation in homothetic coordinates}

We write the conformal scalar wave equation
\eqref{eq:conformal-semilinear} in homothetic coordinates.  The
conformal metric reads
\[ g = -(1-\rho^2) d\tau^2 - 2 \rho\, d\rho\, d\tau + d\rho^2 + \rho^2 d\sigma^2. \]
This metric has constant scalar curvature $\mathcal R[g]=12$.  In
spherical coordinates $(\tau,\rho,\sigma)$, the semilinear wave
equation for the conformal scalar $\psi=t \phi$ becomes
\begin{equation}\label{eq:homothetic_wave}
\begin{aligned}
  -\psi_{\tau\tau}
  &-2\rho\,\psi_{\tau\rho}
  +(1-\rho^2)\,\psi_{\rho\rho}
  -3\,\psi_\tau
  +\left(\frac{2}{\rho}-4\rho\right)\psi_\rho 
  +\frac{1}{\rho^2}\Delta_{\mathbb S^2}\psi
  -2\psi + \Omega^{p-3}\psi^p = 0.
\end{aligned}
\end{equation}

\begin{figure}[t]
  \centering 
 \includegraphics[width=0.25\textwidth]{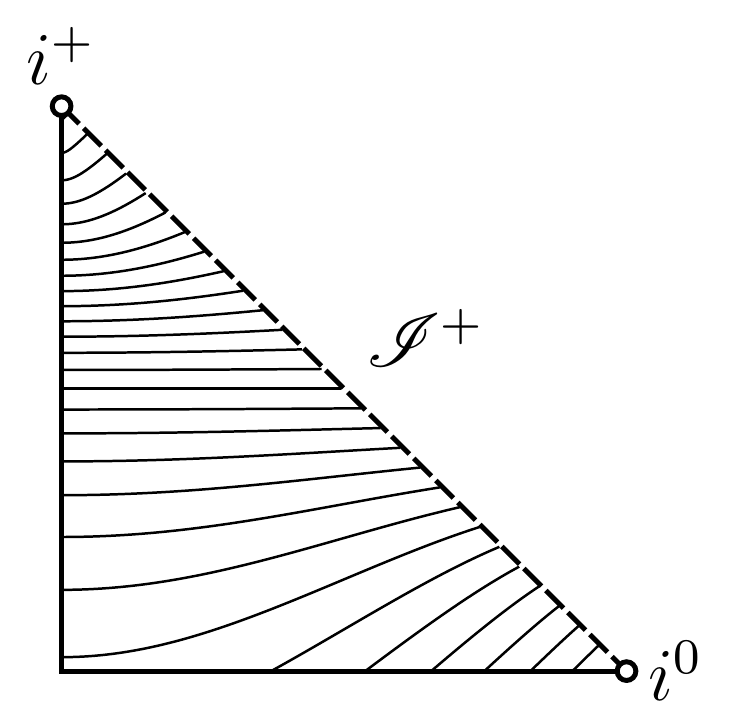}\hspace{1.4cm}
 \includegraphics[width=0.25\textwidth]{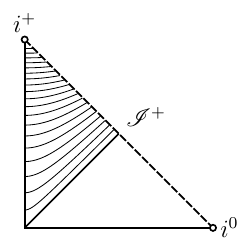}
  \caption{Penrose diagrams of Minkowski spacetime for $t>0$ showing the stationary
  hyperboloidal \eqref{eq:h_of_r} and homothetic hyperboloidal
  \eqref{eq:homothetic_inverse_map} foliations.}
 \label{fig:penrose_diagrams}
\end{figure}

Each $\tau=\mathrm{const}$ slice reaches $\scri$ at the cut
\[
  u\big|_{\scri}=R\,e^\tau.
\]
In particular, the slices foliate the portion of future null infinity with $u\in(0,\infty)$,
corresponding to the region $t>r$ in the Penrose diagram (see Fig.~\ref{fig:penrose_diagrams}).

\subsection{Energy balance}
To derive the energy balance law, we multiply
\eqref{eq:homothetic_wave} by $\rho^2$ and write the
$(\tau,\rho)$-part as a divergence,
\begin{equation}
\partial_\tau\!\left(-\rho^2(\psi_\tau+\rho\,\psi_\rho)\right)
+\partial_\rho\!\left(\rho^2(1-\rho^2)\psi_\rho-\rho^3\psi_\tau\right)
+\Delta_{\mathbb S^2}\psi
-2\rho^2\psi
+\rho^2\Omega^{p-3}\psi^p
=0.
\end{equation}
Multiply the above equation by $-\psi_\tau$. Using only product rules and the identities
\begin{align}
\rho^2\psi_\tau\psi_{\tau\tau}&=\partial_\tau\left(\frac{\rho^2}{2}\psi_\tau^2\right), 
\label{eq:id-time}\\
2\rho^3\psi_\tau\psi_{\tau\rho}+3\rho^2\psi_\tau^2
&=\partial_\rho\left(\rho^3\psi_\tau^2\right),
\label{eq:id-cross}\\
-\rho^2(1-\rho^2)\psi_\tau\psi_{\rho\rho}-(2\rho-4\rho^3)\psi_\tau\psi_\rho
&=-\partial_\rho\left(\rho^2(1-\rho^2)\psi_\tau\psi_\rho\right)
+\rho^2(1-\rho^2)\psi_{\tau\rho}\psi_\rho,
\label{eq:id-radial}\\
\rho^2(1-\rho^2)\psi_{\tau\rho}\psi_\rho
&=\partial_\tau\left(\frac{\rho^2(1-\rho^2)}{2}\psi_\rho^2\right),
\label{eq:id-radial2}\\
-\rho^2 \Omega^{p-3} \psi^p \psi_\tau 
&= -\partial_\tau \left(\frac{\rho^2}{p+1}\Omega^{p-3}\psi^{p+1}\right) + \frac{\rho^2}{p+1} \left(\partial_\tau \Omega^{p-3}\right) \psi^{p+1},
\end{align}
together with the standard sphere identity
\begin{equation}
-\psi_\tau\,\Delta_{\mathbb S^2}\psi
=
\partial_\tau\left(\frac{1}{2}\lvert\nabla_{\mathbb S^2}\psi\rvert^2\right)
+\nabla_{\mathbb S^2}\cdot\left(-\psi_\tau\,\nabla_{\mathbb S^2}\psi\right),
\label{eq:id-sphere}
\end{equation}
and the potential identity $2\rho^2\psi\psi_\tau=\partial_\tau(\rho^2\psi^2)$, we obtain the local
conservation law with a right-hand side
\begin{equation}
\partial_\tau e + \partial_\rho j_\rho + \nabla_{\mathbb S^2}\cdot j_{\mathbb S^2}=s,
\label{eq:homothetic-local-balance}
\end{equation}
where the conformal Noether density and fluxes are
\begin{align}
e(\tau,\rho,\sigma)
&=
\frac{\rho^2}{2}\left(\psi_\tau^{2}+(1-\rho^2)\psi_\rho^{2}\right)
+\frac{1}{2}\lvert\nabla_{\mathbb S^2}\psi\rvert^{2}
+\rho^2\psi^{2}
-\frac{\rho^2}{p+1}\,\Omega^{p-3}\,\psi^{p+1},
\label{eq:homothetic-e}
\\
j_\rho(\tau,\rho,\sigma)
&=
\rho^3\,\psi_\tau^{2}
-\rho^2(1-\rho^2)\,\psi_\tau\,\psi_\rho,
\label{eq:homothetic-jrho}
\\
j_{\mathbb S^2}(\tau,\rho,\sigma)
&=
-\psi_\tau\,\nabla_{\mathbb S^2}\psi .
\label{eq:homothetic-js2}
\end{align}
In contrast to the stationary case, for \(p\ne3\) the balance law
contains an additional term due to the explicit time dependence of
the conformal factor.  The product-rule identity gives
\begin{equation}
\partial_\tau e+\partial_\rho j_\rho+
\nabla_{\mathbb S^2}\cdot j_{\mathbb S^2}
+
\frac{\rho^2}{p+1}
(\partial_\tau\Omega^{p-3})\psi^{p+1}
=0 .
\end{equation}
Equivalently, if the balance law is written with a right-hand side
as in \eqref{eq:homothetic-local-balance}, then
\begin{equation}\label{eq:homothetic_source}
s(\tau,\rho,\sigma)
=
-\frac{\rho^2}{p+1}\,
(\partial_\tau\Omega^{p-3})\,\psi^{p+1}
=
\frac{p-3}{p+1}\,
\rho^2\,\Omega^{p-3}\,\psi^{p+1}.
\end{equation}
Since \(p\) is odd, \(p+1\) is even, and therefore
\(\psi^{p+1}\ge0\).  Thus for \(p>3\) this right-hand-side source
is nonnegative.  It vanishes for \(p=3\), where the conformal
equation is invariant under dilatations.
Integrating \eqref{eq:homothetic-local-balance} over $[0,1]\times\mathbb S^2$ with the usual
assumptions, we obtain the global balance law
\begin{equation}
\frac{d}{d\tau}E(\tau)
=
-\int_{\mathbb S^2}\psi_\tau(\tau,1,\sigma)^2\,d\sigma
+
\frac{p-3}{p+1}\int_0^1\int_{\mathbb S^2}
\rho^2\,\Omega^{p-3}(\tau,\rho)\,
\psi(\tau,\rho,\sigma)^{p+1}\,d\sigma\,d\rho .
\end{equation}
For $\tau_1<\tau_2$, the integrated balance identity reads
\begin{equation}\label{eq:homothetic-integrated-balance}
\begin{aligned}
E(\tau_2)-E(\tau_1) =
&-\int_{\tau_1}^{\tau_2}\int_{\mathbb S^2}
\psi_\tau(\tau,1,\sigma)^2\,d\sigma\,d\tau \\
&+\frac{p-3}{p+1}
\int_{\tau_1}^{\tau_2}\int_0^1\int_{\mathbb S^2}
\rho^2\,\Omega^{p-3}(\tau,\rho)\,
\psi(\tau,\rho,\sigma)^{p+1}
\,d\sigma\,d\rho\,d\tau.
\end{aligned}
\end{equation}
The first term on the right-hand side of
Eq.~\eqref{eq:homothetic-integrated-balance} is the outgoing flux
through future null infinity.  For odd $p>3$, the second term is a
nonnegative bulk contribution arising from the explicit
$\tau$-dependence of the conformal nonlinear potential.  Therefore, the
homothetic balance quantity is not generally monotone. The outgoing flux through $\scri$ decreases energy, but the time dependence of the focusing potential gives a competing nonnegative contribution.


\section{Uniform exponential decay rates in homothetic coordinates}
\label{sec:rates}

The homothetic construction is useful for tail computations because it
is adapted to the self-similar structure of late-time asymptotics.  In
this section we explain this point at the level of decay rates.  We
first review the spherical case, where rigorous small-data asymptotics
identify the different timelike and null-infinity rates.  We then
discuss the higher-multipole rates proposed in \cite{Rinne:2025} and
the role of origin regularity.  These two examples motivate a
conditional homothetic-coordinate calculation showing how an assumed
self-similar radiative tail approaches a fixed radial profile
with a uniform exponential decay rate.

We use the following notation throughout this section.  For a
multipole \(\phi_{\ell m}\) of the physical scalar field, define
the radiative mode $\chi_{\ell m}(t,r):=r\,\phi_{\ell m}(t,r)$.
In the homothetic formulation the conformal unknown is instead
\[
        \psi_{\ell m}:=t\,\phi_{\ell m}
        =\frac{\chi_{\ell m}}{\rho},
        \qquad
        \rho=\frac{r}{t}.
\]
Thus \(\psi_{\ell m}\) and \(\chi_{\ell m}\) agree at future null
infinity, where \(\rho=1\), but not in the interior of the
homothetic domain.  The tail measured
at \(\scri\) is the tail of the radiative field, but the field
evolved numerically is the conformal unknown.

\subsection{Spherical rates}

The long-time solution of semilinear wave equations of the type
\eqref{eq:standard_wave} for small data decays polynomially with
rates that depend on the asymptotic regime.  In spherical symmetry,
rigorous asymptotic results include a uniform estimate in the future
timelike cone \(t\ge r\) of the form
\cite{Szpak:2008jv,Szpak:2009}
\begin{equation}\label{eq:SzpakBound}
  |\phi(t,r)| \le
  \frac{C}{(1+t+r)\,(1+t-r)^{p-2}},
\end{equation}
where \(C\) depends on the data and on \(p\).  This single estimate
encodes both the faster timelike decay and the slower null-infinity
decay.

First consider the timelike interior limit, \(t\to\infty\) with
\(r\) fixed.  Then \(1+t+r\sim t\) and \(1+t-r\sim t\), so
\eqref{eq:SzpakBound} gives $|\phi(t,r)|\lesssim t^{-(p-1)}$.
For generic spherical data this rate is sharp, and the leading
coefficient can be computed explicitly
\cite{Szpak:2008jv,Szpak:2009}. At future null infinity the relevant quantity is the
radiation field $\chi(t,r):=r\,\phi(t,r)$.
Multiplying \eqref{eq:SzpakBound} by \(r\) gives
\begin{equation}\label{eq:chiBound}
  |\chi(t,r)| \le
  \frac{C\,r}{(1+t+r)\,(1+t-r)^{p-2}} .
\end{equation}
Taking $r\to\infty$ at fixed retarded time $u=t-r$ defines the
radiation field at future null infinity,
\[
  \chi_{\scri}(u):=\lim_{r\to\infty}\chi(u+r,r).
\]
The late-time limit along $\scri$ is then $u\to\infty$.  Since
$t+r\sim2r$ in the limit defining the radiation field,
Eq.~\eqref{eq:chiBound} implies
$|\chi_{\scri}(u)|\lesssim u^{-(p-2)}$ at late retarded time.
Therefore, the radiation field at null infinity decays one power more
slowly than the physical field along finite-radius timelike worldlines
\cite{Bizon:2008zd,Szpak:2009,Szpak:2010}.  This split is the source
of the steepening radial profiles seen in long stationary
hyperboloidal evolutions on compactified grids
\cite{Zenginoglu:2008wc}.

Now consider the homothetic limit, \(t\to\infty\) with
\(\rho=r/t\) fixed.  Let $q_0:=p-2$.
Using \eqref{eq:homothetic_chars}
and the bound
\[
        \frac{r}{1+t+r}\le \frac{\rho}{1+\rho},
\]
equation \eqref{eq:chiBound} gives, for \(0\le\rho\le1\),
\begin{equation}\label{eq:chiTauBound}
  |\chi(\tau,\rho)|
  \lesssim
  e^{-q_0\tau}\,G_0(\rho),
  \qquad
  G_0(\rho):=(2R)^{-q_0}\rho(1+\rho)^{q_0-1}.
\end{equation}
The prefactor \(G_0\) is smooth on the compactified interval
\([0,1]\). 
For every fixed \(0<\rho\le1\), the conformal unknown
\(\psi=\chi/\rho\) has the same exponential rate.  The quotient \(G_0(\rho)/\rho\) is regular at the origin, so the
homothetic profile remains smooth across the ball.

The faster timelike decay is not a different decay law at a fixed
homothetic radius.  Along a fixed physical radius $r>0$, let
$t\to\infty$.  This implies
\begin{equation}\label{eq:homothetic_limit}
  \rho=r/t\to0,\qquad u\sim t,\qquad \tau\to\infty.
\end{equation}
In that limit the
regularity factor \(G_0(\rho)\sim \rho\) supplies one additional
factor of \(t^{-1}\), giving the fixed-radius rate
\(t^{-(p-1)}\).  This is the basic mechanism by which
homothetic coordinates produce the same decay at every compactified radius.

\subsection{Higher multipole rates and origin regularity}
\label{sec:higher-multipole-rates}

The spherical rates above follow from rigorous small-data asymptotics
\cite{Szpak:2008jv,Szpak:2009}.  For higher multipoles, there is not
yet an analogous theorem for the semilinear problem considered here.
Nevertheless, the rates proposed in \cite{Rinne:2025} have a simple
coordinate interpretation in homothetic coordinates. Let
\[
        \phi(t,r,\sigma)
        =
        \sum_{\ell,m}\phi_{\ell m}(t,r)Y_{\ell m}(\sigma),
\]
and define the radiative modes $\chi_{\ell m}(t,r):=r\,\phi_{\ell m}(t,r)$.
In homothetic variables the evolved conformal unknown is
\[
        \psi_{\ell m}=t\phi_{\ell m}
        =\frac{\chi_{\ell m}}{\rho}.
\]
As $r\to0$ at fixed $t$, smoothness at the origin gives the
standard expansions:
\begin{equation}\label{eq:originReg}
        \phi_{\ell m}=\O(r^\ell),
        \qquad
        \chi_{\ell m}=\O(r^{\ell+1}),
        \qquad
        \psi_{\ell m}=\O(\rho^\ell).
\end{equation}
Rinne's numerical study of nonlinear tails in \(3+1\) dimensions
reports that the late-time decay at \(\scri\) and at fixed finite
radius is described by the exponents
\begin{equation}\label{eq:RinneExponents}
  \tilde q_\ell=\max(p-2,\ell+1)
  \quad \text{at }\scri,
  \qquad
  q_\ell=\max(\ell+p-1,2\ell+2)
  \quad \text{at fixed }r.
\end{equation}
See Table 2 and Conjecture 1 of \cite{Rinne:2025}.  The two formulas in \eqref{eq:RinneExponents}
differ by $q_\ell-\tilde q_\ell=\ell+1$.
This difference is naturally produced if the radiative mode has a
self-similar null-infinity tail whose profile is compatible with
origin regularity.  More precisely, suppose heuristically that
\begin{equation}\label{eq:similarityAnsatz}
        \chi_{\ell m}(t,r)
        \sim
        u^{-\tilde q_\ell}F_{\ell m}(\rho),
        \qquad
        \rho=\frac{r}{t},
\end{equation}
with
\[
        F_{\ell m}(\rho)\sim \rho^{\ell+1}
        \qquad
        \text{as }\rho\to0,
\]
as required by \eqref{eq:originReg}.  At fixed homothetic radius the
exponent should then be the null-infinity exponent \(\tilde q_\ell\).
Along a fixed physical radius \(r>0\), however, letting
\(t\to\infty\) gives \(\rho=r/t\to0\), \(u\sim t\), and
\(\tau\to\infty\).  The origin factor
\(F_{\ell m}(\rho)\sim\rho^{\ell+1}\) then supplies
\(\ell+1\) additional powers of \(t^{-1}\).  

\subsection{Homothetic coordinate mechanism for tail decay}
\label{sec:homothetic-coordinate-mechanism}

The spherical discussion and the higher-multipole ansatz point to the
same coordinate mechanism.  Below we assume that a leading
asymptotic representation exists and show how it transforms under the
homothetic change of variables.

Fix a mode \((\ell,m)\) from the decomposition introduced above.
Since \(m\) plays no role in the following coordinate argument, we
suppress it throughout this subsection and write
\(\phi_\ell,\chi_\ell,\psi_\ell\).  We use the homothetic coordinate
relations already defined in
Eqs.~\eqref{eq:homothetic_compact_spatial}--\eqref{eq:homothetic_chars}.

Suppose that, for some $q>0$ and $\delta>0$, the radiative mode
has the leading asymptotic representation
\begin{equation}\label{eq:self-similar-tail-assumption}
  \chi_\ell(u,\rho)
  =
  u^{-q}F_\ell(\rho)
  +
  \mathcal R_\ell(u,\rho),
\end{equation}
where we assume that the the leading profile has the origin-regular form
\begin{equation}\label{eq:origin-compatible-tail-profile}
  F_\ell(\rho)
  =
  \rho^{\ell+1}K_\ell(\rho),
  \qquad
  K_\ell\in C^\infty([0,1]),
\end{equation}
and where, as $u\to\infty$, uniformly in $\rho\in[0,1]$, the remainder is assumed to satisfy
\begin{equation}\label{eq:self-similar-tail-remainder}
  \mathcal R_\ell(u,\rho)
  =
  \O\!\left(u^{-q-\delta}\rho^{\ell+1}\right),
  \qquad
  u\partial_u\mathcal R_\ell(u,\rho)
  =
  \O\!\left(u^{-q-\delta}\rho^{\ell+1}\right).
\end{equation}
Writing the remainder in homothetic coordinates as
$\mathcal R_\ell(\tau,\rho)$ and substituting
$u=\frac{2Re^\tau}{1+\rho}$ from Eq.~\eqref{eq:homothetic_chars} into
Eq.~\eqref{eq:self-similar-tail-assumption} gives
\[
  \chi_\ell(\tau,\rho)
  =e^{-q\tau}(2R)^{-q}(1+\rho)^qF_\ell(\rho)
  +\mathcal R_\ell(\tau,\rho).
\]
We therefore define the homothetic radiative profile
\begin{equation}\label{eq:homothetic-radiative-profile}
  P_\ell(\rho)
  :=
  (2R)^{-q}(1+\rho)^qF_\ell(\rho)
  =
  (2R)^{-q}(1+\rho)^q\rho^{\ell+1}K_\ell(\rho).
\end{equation}
Equation~\eqref{eq:self-similar-tail-remainder} then gives, uniformly in
$\rho\in[0,1]$,
\[
  \mathcal R_\ell(\tau,\rho)
  =\O\!\left(e^{-(q+\delta)\tau}\rho^{\ell+1}\right),
  \qquad
  \partial_\tau\mathcal R_\ell(\tau,\rho)
  =\O\!\left(e^{-(q+\delta)\tau}\rho^{\ell+1}\right).
\]
Here the derivative relation at fixed $\rho$ is
$(\partial_\tau)_\rho=u(\partial_u)_\rho$.  As
$\tau\to\infty$, uniformly in $\rho\in[0,1]$,
\begin{equation}\label{eq:chi-homogeneous-tail}
  \chi_\ell(\tau,\rho)
  =
  e^{-q\tau}P_\ell(\rho)
  +
  \O\!\left(e^{-(q+\delta)\tau}\rho^{\ell+1}\right).
\end{equation}

Dividing Eq.~\eqref{eq:chi-homogeneous-tail} by $\rho$, define
\begin{equation}\label{eq:homothetic-conformal-profile}
  Q_\ell(\rho)
  :=
  \frac{P_\ell(\rho)}{\rho}
  =
  (2R)^{-q}(1+\rho)^q\rho^\ell K_\ell(\rho).
\end{equation}
The right-hand side shows that \(Q_\ell\) extends smoothly to
\(\rho=0\).  Using $\psi_\ell=\chi_\ell/\rho$ then gives, as
$\tau\to\infty$, uniformly in $\rho\in[0,1]$,
\begin{equation}\label{eq:psi-homogeneous-tail}
  \psi_\ell(\tau,\rho)
  =
  e^{-q\tau}Q_\ell(\rho)
  +
  \O\!\left(e^{-(q+\delta)\tau}\rho^\ell\right).
\end{equation}

For every fixed \(0<\rho\le1\) such that \(P_\ell(\rho)\ne0\), we write
\[
  \chi_\ell(\tau,\rho)
  =
  e^{-q\tau}P_\ell(\rho)
  \bigl(1+\varepsilon_\ell(\tau,\rho)\bigr),
  \qquad
  \varepsilon_\ell
  :=
  \frac{e^{q\tau}}{P_\ell(\rho)}
  \mathcal R_\ell(\tau,\rho).
\]
The remainder and its derivative imply
\(\varepsilon_\ell=\O(e^{-\delta\tau})\) and
\(\partial_\tau\varepsilon_\ell=\O(e^{-\delta\tau})\).  Therefore
\begin{equation}\label{eq:local-homothetic-rate}
  -\partial_\tau\ln|\chi_\ell(\tau,\rho)|
  =
  q-\frac{\partial_\tau\varepsilon_\ell}{1+\varepsilon_\ell}
  =
  q+\O(e^{-\delta\tau})
  \longrightarrow q
  \qquad
  \text{as }\tau\to\infty.
\end{equation}
The derivative estimate in
\eqref{eq:self-similar-tail-remainder} is essential for this
conclusion.  The statement is pointwise at each fixed homothetic radius
where the leading profile is nonzero.

The fixed-radius behavior is obtained by a different limiting
process.  Along a fixed physical radius $r>0$, let
$t\to\infty$ which implies \eqref{eq:homothetic_limit}.
Equation~\eqref{eq:origin-compatible-tail-profile}
then gives
\[
  F_\ell(r/t)
  =
  \left(\frac{r}{t}\right)^{\ell+1}K_\ell(r/t).
\]
The uniform expansion can therefore be evaluated along this
fixed-radius curve:
\[
  \chi_\ell(t,r)
  =
  (t-r)^{-q}
  \left(\frac{r}{t}\right)^{\ell+1}K_\ell(r/t)
  +
  \O\!\left(
    (t-r)^{-q-\delta}
    \left(\frac{r}{t}\right)^{\ell+1}
  \right).
\]
Provided \(K_\ell(0)\ne0\), this implies
\begin{equation}\label{eq:fixed-radius-from-homogeneous-tail}
  \begin{aligned}
    \chi_\ell(t,r)
    &\sim
    r^{\ell+1}K_\ell(0)t^{-(q+\ell+1)},\\
    \phi_\ell(t,r)
    =
    \frac{\chi_\ell(t,r)}{r}
    &\sim
    r^\ell K_\ell(0)t^{-(q+\ell+1)}
  \end{aligned}
  \qquad
  \text{as }t\to\infty.
\end{equation}
The additional \(\ell+1\) powers arise from the origin factor
\(\rho^{\ell+1}\).
The difference between the null-infinity and fixed-radius rates
reflects two different limiting processes.  If \(K_\ell(0)=0\), the
displayed leading fixed-radius coefficient vanishes; determining the
decay then requires a fuller joint late-time and origin expansion,
whose first nonvanishing term controls the result.

For the spherical case, \(q=q_0=p-2\) and
\(F_0(\rho)=\rho K_0(\rho)\).  At each fixed
\(0<\rho\le1\) where \(P_0(\rho)\ne0\), the coordinate calculation
gives the exponential rate \(e^{-(p-2)\tau}\).  If
\(K_0(0)\ne0\), \eqref{eq:fixed-radius-from-homogeneous-tail} gives
the fixed-radius physical-field decay \(t^{-(p-1)}\).

For the heuristic higher-multipole ansatz
\eqref{eq:similarityAnsatz}, take \(q=\tilde q_\ell\).  With the
\(m\)-index still suppressed, the homothetic behavior is
schematically
\[
  \chi_\ell(\tau,\rho)
  \sim
  e^{-\tilde q_\ell\tau}P_\ell(\rho),
  \qquad
  \psi_\ell(\tau,\rho)
  \sim
  e^{-\tilde q_\ell\tau}Q_\ell(\rho).
\]
Along a fixed physical radius \(r>0\), let \(t\to\infty\), so that
\(\rho=r/t\to0\), \(u\sim t\), and \(\tau\to\infty\).  Provided
\(K_\ell(0)\ne0\), the origin-regularity factor supplies
\(\ell+1\) additional powers of \(t^{-1}\).  Using
\eqref{eq:RinneExponents},
\[
  \tilde q_\ell+\ell+1
  =
  \max(p-2,\ell+1)+\ell+1
  =
  \max(\ell+p-1,2\ell+2)
  =
  q_\ell,
\]
which reproduces the fixed-radius exponent proposed in
\cite{Rinne:2025} as a kinematic consequence of the assumed tail and
its origin-regularity factor.


\section{Numerical results}
\label{sec:results}

We now test our main claim that homothetic coordinates keep
the late-time radial profile resolved and produce the same exponential
decay rate at every fixed compactified radius, in contrast to stationary
hyperboloidal coordinates that develop an increasingly sharp transition
near $\scri$.  We first
validate the two discretized formulations of stationary and homothetic equations in
Sec.~\ref{sec:results-validation}: the convergence test in
Sec.~\ref{sec:results-convergence} checks spectral accuracy; the energy-balance
diagnostic in Sec.~\ref{sec:results-energy-balance} verifies the
implemented conformal balance laws in controlled linear evolutions.  We
then turn to tail extraction in Sec.~\ref{sec:tail-analysis}, where we compare stationary profile
steepening with the fixed
homothetic tail profile. Sec.~\ref{sec:computational-efficiency}
uses this benchmark to quantify the computational advantage of the
homothetic formulation. Sec.~\ref{sec:noncompact-scri-data} records
the separate observation that initial data with non-negligible support
at $\scri$ can decay more slowly than compactly supported or rapidly
decaying data. The higher-mode survey in
Sec.~\ref{sec:higher-mode-survey} tests the expected null-infinity
rates for several powers and angular modes. Finally,
Sec.~\ref{sec:non-generic-counterexample} presents numerical evidence for a nongeneric cubic
monopole tail.

\subsection{Validation and accuracy}
\label{sec:results-validation}

All stationary evolutions presented below use the
time-translated hyperboloids and Poincar\'e-ball compactification of
Sec.~\ref{sec:stationary-hyperboloids}.  The homothetic evolutions use
the coordinates of Sec.~\ref{sec:homothetic}.

\begin{figure}[t]
  \centering 
 \includegraphics[width=\textwidth]{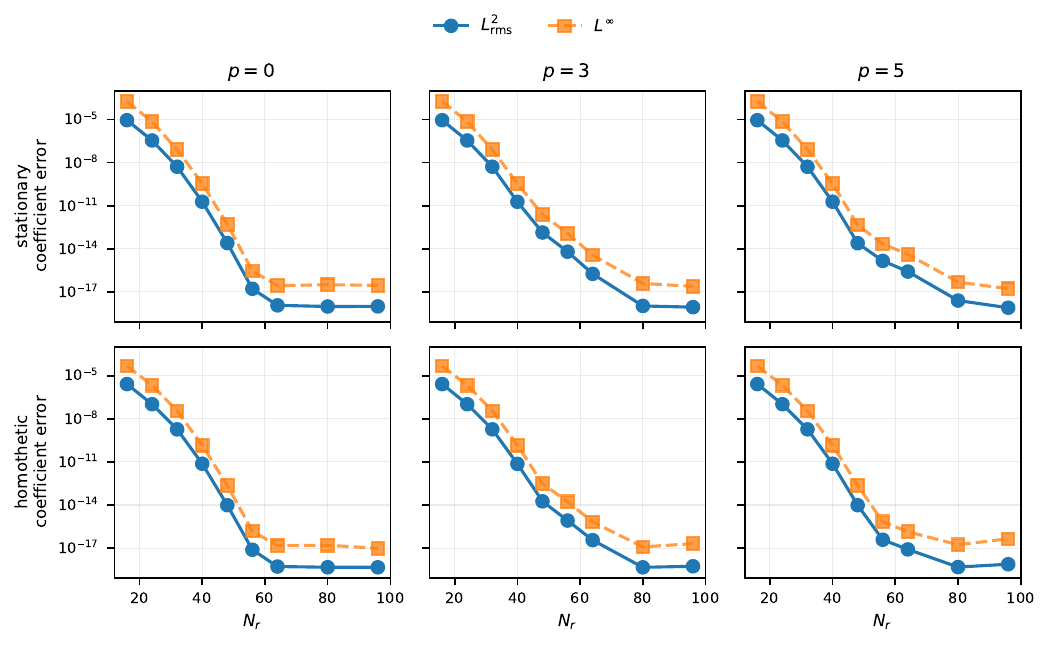}
  \caption{Coefficient-space self-convergence for the stationary
  hyperboloidal formulation (top row) and the homothetic formulation
  (bottom row), for $p=0,3,5$.  The plotted quantities are the RMS
  coefficient error $E_{2,\mathrm{rms}}$ and the absolute maximum
  coefficient error $E_\infty$ at $\tau_{\mathrm{end}}=2$, measured
  against the $N_r=128$ reference solution.  The test uses
  $\ell=1$, $m=0$, $A=100$, $\sigma=0.1$,
  $(N_\theta,N_\vphi)=(16,4)$, $\Delta\tau=0.005$, and
  $N_r=16,24,32,40,48,56,64,80,96$.}
 \label{fig:convergence}
\end{figure}

\subsubsection{Convergence tests}
\label{sec:results-convergence}
Details of the numerical implementation are discussed in
App.~\ref{app:numerics}. Here, we verify that the radial
discretization is spectrally accurate by performing a convergence test
on the final-time numerical solution for both the stationary and homothetic formulations.  We use
the same regular,
axisymmetric, origin-centered initial data in all runs,
\begin{equation}
  \psi(0,\rho,\theta,\varphi)=0,\qquad
  \psi_\tau(0,\rho,\theta,\varphi)
  = A\,\rho^\ell \exp[-(\rho/\sigma)^2]Y_{\ell 0}(\theta,\varphi),
\end{equation}
with $\ell=1$, $A=100$, and $\sigma=0.1$. We
run the linear equation ($p=0$), the conformally invariant cubic
equation ($p=3$), and the quintic equation ($p=5$) to
$\tau_{\mathrm{end}}=2$ with time step $\Delta\tau=0.005$ and fixed
angular resolution $(N_\theta,N_\vphi)=(16,4)$.  The radial resolutions
are $N_r=16,24,32,40,48,56,64,80,96$, compared against a reference
computation with $N_r=128$.  We set Dedalus solver tolerances to $10^{-15}$.

The error is measured in coefficient space, without
interpolation.  For each run with $N_r<128$, we truncate the reference
coefficient array to the same size as the $N_r$ array.  Denoting this
truncated array by $c_{128}^{(N_r)}$, we compute
\begin{equation}
  E_{2,\mathrm{rms}}(N_r)
  =\frac{\left\|c_{N_r}-c_{128}^{(N_r)}\right\|_2}
    {\sqrt{n_{N_r}}},
  \qquad
  E_\infty(N_r)
  =\left\|c_{N_r}-c_{128}^{(N_r)}\right\|_\infty,
\end{equation}
where $n_{N_r}$ is the number of coefficients in the $N_r$ array.

Figure~\ref{fig:convergence} shows spectral convergence in both
coordinate systems and for all three powers.  The error decreases by
many orders of magnitude as $N_r$ is increased and then reaches the
reference/roundoff floor.  For the runs shown, the smallest RMS
coefficient errors are approximately $10^{-18}$ in the stationary
formulation and $5\times10^{-19}$ in the homothetic formulation, with
maximum coefficient errors of order $10^{-17}$.  Thus, for these
regular initial data, both implementations exhibit spectral
convergence against the reference solution, with an approximately
exponential error decrease before reaching machine precision.

\subsubsection{Energy-balance diagnostics}
\label{sec:results-energy-balance}

\begin{figure}[t]
  \centering
  \includegraphics[width=\textwidth]{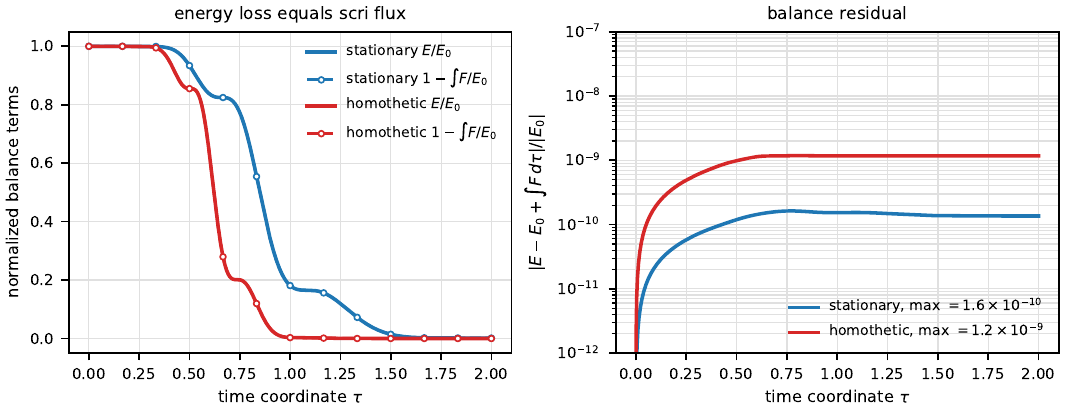}
  \caption{Conformal energy-balance for short linear evolutions.
  Left: normalized energy and the integrated flux prediction for
  stationary and homothetic evolutions.  Right: relative energy
  residual.}
  \label{fig:energy-balance-diagnostics}
\end{figure}

The integrated identities \eqref{eq:integrated-energy-decay-updated}
and \eqref{eq:homothetic-integrated-balance} provide consistency checks
for the numerical evolutions.  For the energy balance, we set the
nonlinear term to zero and evolve the linear equation in both
coordinate systems.  We evolve the
same regular $\ell=2$, $m=1$ pulse in the stationary and homothetic
formulations, as in \cite{Rinne:2025}, with
\[
  \psi(0,\rho,\theta,\varphi)=0,\qquad
  \psi_\tau(0,\rho,\theta,\varphi)
  =
  \rho^2 e^{-(\rho/\sigma)^2}Y_{21}(\theta,\varphi),
  \qquad
  \sigma=0.2 .
\]
As in the convergence test, the Gaussian is centered at the origin so
that the radial factor multiplying $\rho^\ell Y_{\ell m}$ is a smooth
function of $\rho^2$.
The runs use
$(N_r,N_\theta,N_\varphi)=(64,8,12)$. 
Since the balance residual compares the evolved conformal
energy with a separately time-integrated boundary flux, it is
especially sensitive to time-discretization and flux-quadrature error.
We therefore use $\Delta\tau=10^{-4}$ for this diagnostic, smaller
than the timesteps needed in the tail runs, and record the flux every
two time steps.  The conformal density and scri flux are evaluated from
inside the time-stepping code. The stored flux is
then integrated in time by Simpson quadrature.
Figure~\ref{fig:energy-balance-diagnostics} compares the numerical energy $E(\tau)$ with the
prediction
$E(0)-\int_0^\tau F(s)\,ds$, where
$F=\int_{\scri}\psi_\tau^2\,d\sigma$.  The maximum relative residual
$|E(\tau)-E(0)+\int_0^\tau F(s)\,ds|/|E(0)|$ is
$1.63\times10^{-10}$ in the stationary run and
$1.18\times10^{-9}$ in the homothetic run.  Note that the residual is not a
measure of the spectral spatial error alone. It also contains the
Runge--Kutta time-discretization error, interpolation error in the
diagnostic extraction at $\scri$, and the time quadrature error in the
recorded flux integral.  Within this accuracy, the observed change in
the energy is accounted for by the outgoing flux through
$\scri$.

\subsection{Stationary versus homothetic tail extraction}
\label{sec:tail-analysis}
\label{sec:stationary-homothetic-tail-comparison}

The analytic discussion in Sec.~\ref{sec:rates} suggests that the tail has the form
  \(e^{-\tilde q_\ell\tau}P_\ell(\rho)\) in the homothetic formulation, so the same exponential
rate should be measured at every fixed compactified radius
\(\rho\in(0,1]\). In contrast, in stationary hyperboloidal coordinates, the compactified
grid contains the slower radiative
decay at \(\scri\) and the faster decay associated with finite-radius
observers, producing increasingly steep radial profiles at late times.
In this section, we compare these properties numerically. For each run we record
spherical-harmonic coefficients of the conformally rescaled field at fixed extraction radii.
For example, at \(\scri\), $a_{\ell m}(\tau):=\psi_{\ell m}(\tau,\rho=1)$.
This is also the radiative coefficient at future null infinity:
\(\psi_{\ell m}=\chi_{\ell m}/\rho\), so
\(\psi_{\ell m}=\chi_{\ell m}\) at \(\rho=1\).
In homothetic time the expected physical power laws become
exponentials, so we use the corresponding
local logarithmic rate at fixed \(\rho\).  This diagnostic is
motivated by Eq.~\eqref{eq:local-homothetic-rate}: since
$\chi_{\ell m}=\rho\psi_{\ell m}$, at every fixed extraction radius
where the leading profile is nonzero $-\partial_\tau\ln|\psi_{\ell m}|\longrightarrow q$.

\begin{figure}[t]
  \centering
  \includegraphics[width=\textwidth]{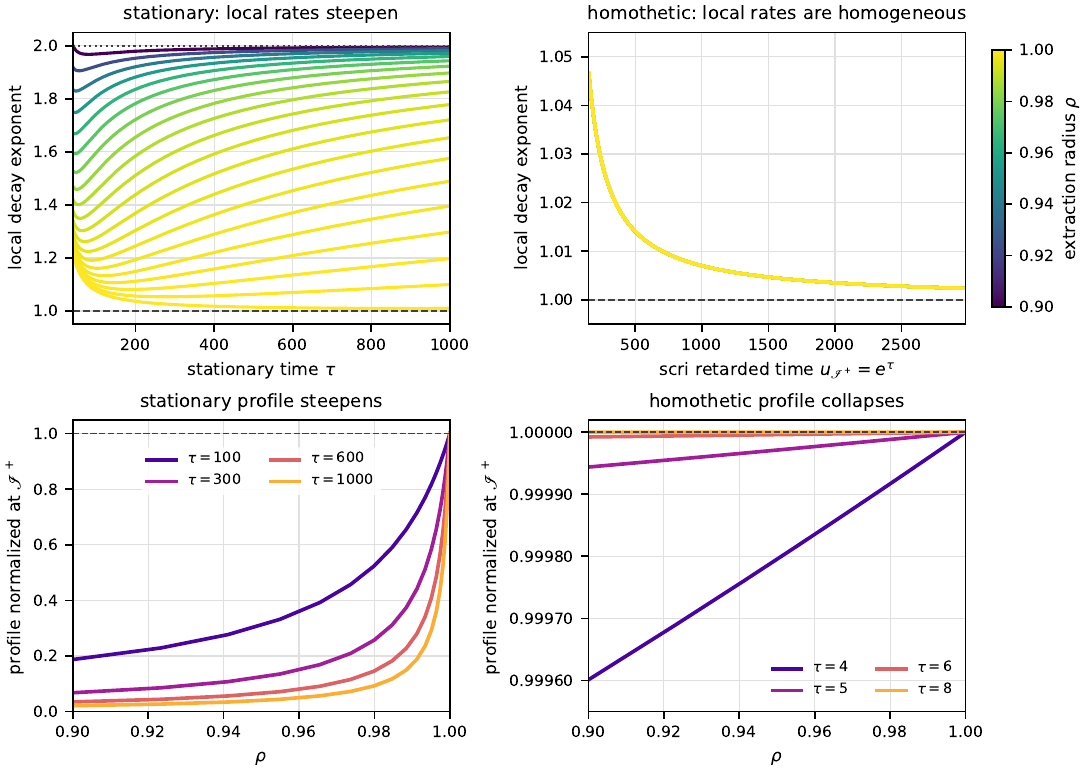}
  \caption{Uniform exponential decay rate of the cubic monopole in
  homothetic coordinates, compared with stationary hyperboloidal
  coordinates. Top left: stationary local power index
  $-d\ln |a_{00}|/d\ln \tau$ at 20 extraction radii clustered in
  \(0.9\le\rho\le1\), shown over the stationary interval
  \(\tau\in[40,1000]\).  Top right: the homothetic local index
  $-d\ln |a_{00}|/d\tau$, plotted against the retarded time
  \(u_{\scri}=e^\tau\) of the corresponding cut of \(\scri\), over
  $\tau\in[5,8]$ which roughly corresponds to \([150,3000]\) as a Killing time interval.  Bottom
  left: stationary profiles
  normalized at \(\scri\), showing the
  steepening transition toward $\scri$.  Bottom right: homothetic
  profiles normalized at \(\scri\),
  showing a fixed radial shape with a uniform decay exponent
  across the grid.  At interior extraction radii, the local retarded
  time differs from the cut time used on the horizontal axis:
  in stationary coordinates \(u=\tau+(1-\rho)/(1+\rho)\); in
  homothetic coordinates \(u=2 e^\tau/(1+\rho)\).  The plotted
  variables are common time labels for the extracted family.  
  The two columns use different vertical
  scales to make both the large stationary rate spread and the small
  residual homothetic variation visible. }
  \label{fig:decay-rate-homogeneity}
\end{figure}

Figure~\ref{fig:decay-rate-homogeneity} compares the local decay rates for stationary and homothetic
formulations on the example of the cubic monopole.  We extract the decay rates at 20 radii clustered
in the interval
\(0.9\le\rho\le1\).  In the stationary panel the horizontal coordinate
is the stationary time \(\tau\). At fixed \(\rho\),
\(u=t-r=\tau+R(1-\rho)/(1+\rho)\).  Thus the plotted stationary local
index is \(-d\ln |a_{00}|/d\ln\tau\), which agrees asymptotically
with the retarded-time power index at each fixed \(\rho\).  In the
homothetic panel the horizontal coordinate is the retarded time of
the cut of \(\scri\), \(u_{\scri}=e^\tau\) in the units used here.
Away from \(\scri\), \(u=2e^\tau/(1+\rho)\); this differs by a
\(\rho\)-dependent factor, so \(-d\ln |a_{00}|/d\tau\) is still the
corresponding logarithmic retarded-time exponent at fixed \(\rho\).

In stationary hyperboloidal coordinates, the local power indices
are distinct. Over the plotted window
\(\tau\in[40,1000]\), the median rates range from
\(q_{\rm loc}=1.991\) at \(\rho=0.9\) to \(q_{\rm loc}=1.014\) at
\(\rho=1\) in accordance with the analytic expectation of $q_{\rm finite}=2$ and $q_\scri=1$.  
In homothetic coordinates, the corresponding local rates
agree across all extraction radii to the displayed accuracy:
over \(\tau\in[5,8]\), the
medians over all 20 radii are \(1.010\) to the quoted precision, with
a total spread of only \(6\times10^{-6}\). Note that this homothetic interval corresponds to about \([150,3000]\)
as a Killing time interval. At the common scri retarded time \(u_{\scri}=1000\),
the stationary run gives \(q_{\rm loc}=1.009\) at \(\scri\) and
the homothetic run gives \(q_{\rm loc}=1.007\).
Considering that the initial hypersurfaces in the
two systems are not the same, this agreement is
satisfactory for our purposes.  The homothetic computation is already visibly
more efficient in this example, both because it reaches larger
retarded times with logarithmic time integration and because the
fixed tail profile has lower radial-resolution requirements.
The homothetic run shown here uses \(N_r=48\), compared with
\(N_r=96\) for the stationary run.  Once the solution has reached
\(u_{\scri}\simeq1000\), reaching \(u_{\scri}\simeq3000\) requires only
about one additional unit of homothetic time, since
\(\Delta\tau\approx \ln 3\).  We return to this exponential efficiency in
Sec.~\ref{sec:computational-efficiency}.
The normalized radial profiles on the lower panels of Fig.~\ref{fig:decay-rate-homogeneity} show the same effect.
Stationary profiles steepen toward \(\scri\) caused by the observer-dependent power laws on a
compactified grid.
Homothetic profiles approach a fixed shape with a uniform
exponential decay rate across the grid,
providing numerical evidence for the discussion
in Sec.~\ref{sec:rates}.

\subsection{Computational efficiency}
\label{sec:computational-efficiency}

At every fixed grid point $\rho\in(0,1]$, the radiative mode has the
asymptotic form $\chi_\ell(\tau,\rho)
  \sim e^{-\widetilde q_\ell\tau}P_\ell(\rho)$,
so the decay exponent is the same at all compactified radii and
the radial profile is smooth. At
$\scri$, $u=Re^\tau$, so reaching a retarded time $U$ requires
$N_{\rm stat}\simeq U/\Delta u$ steps in stationary hyperboloidal time,
whereas $N_{\rm hom}\sim\ln(U/R)/\Delta\tau$ in homothetic time. Along a fixed physical radius $r>0$, letting $t\to\infty$ implies \eqref{eq:homothetic_limit}.
The origin-regularity factor $\rho^{\ell+1}$ then recovers the faster
finite-radius decay.  At $\scri$, the radiative mode and conformal
unknown agree.

\begin{figure}[t]
  \centering
  \includegraphics[width=\textwidth]{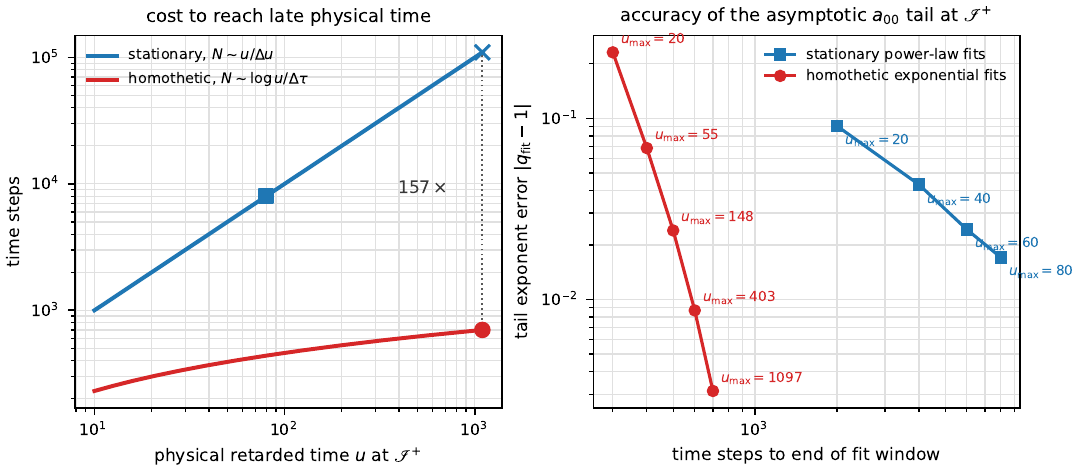}
  \caption{Computational cost and tail-fit accuracy for the cubic
  monopole at $\scri$.  Left: time-step count required to reach a
  retarded time $u$ at $\scri$ for stationary and homothetic
  evolutions using the same timestep,
  $\Delta u=\Delta\tau=0.01$.  The square marks the completed stationary run
  $u=80$, the circle marks the completed homothetic run
  $u=e^7$, and the cross marks the projected stationary step count
  needed to reach the same retarded-time endpoint as the homothetic run.
  Right: accuracy of the fitted asymptotic exponent, measured by
  $|q_{\rm fit}-1|$, as a function of time steps to the end of the
  fit window.  The blue squares use stationary power-law fits in
  $u$-windows; the red circles use homothetic exponential fits.  Labels show the retarded time endpoint $u_{\max}$ of
  each fit.}
  \label{fig:tail-efficiency-benchmark}
\end{figure}

To demonstrate the gain of efficiency, we compare the time-step cost with the
error in the fitted scri exponent for the cubic monopole, whose
expected generic value is $\tilde q_0=1$.  In the stationary run we
fit $\ln |a_{00}|$ against $\ln u$; in the homothetic run we fit $\ln |a_{00}|$ against
$\tau$.  The error measure is $|q_{\rm fit}-1|$.
Figure~\ref{fig:tail-efficiency-benchmark} shows a comparison of numerical time steps for $N_r=96$.
The stationary and homothetic runs use the same timestep,
$\Delta u=\Delta\tau=0.01$, and the same spatial resolution.
The stationary computation
reaches $u=80$ in $8000$ time steps with a fit 
on $u\in[60,80]$ giving $q_{\rm fit}=1.017$.  The
homothetic computation reaches
$\tau=7$, equivalently $u=e^7\simeq1097$, in $700$ time steps and the fit on $\tau\in[6,7]$ gives $q_{\rm fit}=1.003$.
A stationary computation with the same
timestep would require $1.10\times10^5$ steps to reach
$u=e^7$, a factor of $156.7$ more than the homothetic run.  Therefore the
homothetic calculation delivers a more accurate asymptotic tail exponent
for a substantially lower step count.

The homothetic construction is well-suited for tail extraction,
where the solution has entered the self-similar regime described
above.  However, if essential wave dynamics occurs at early times,
then a purely homothetic evolution may not be an efficient
description of that part of the problem. In particular, an extension
of this construction to black-hole spacetimes will likely need a
modified time coordinate that matches an interior stationary horizon-penetrating system to an asymptotically homothetic system,
or matching from an early-time calculation to a tail-adapted
homothetic calculation.  Schwarzschild and Kerr spacetimes introduce the
mass as intrinsic scales and therefore the corresponding
tail problem is not exactly scale invariant.  Extending the efficiency
mechanism to black-hole backgrounds requires a new
construction adapted to those scales, and is left for future work.

\subsection{Non-compact data at \texorpdfstring{$\scri$}{scri}}
\label{sec:noncompact-scri-data}

The tail-rate tests above use rapidly decaying
data that are compactly supported for numerical purposes. The conjecture is valid for such data.
However, the behavior of decay rates is different for hyperboloidal data that have support at the
boundary. To demonstrate this, we set Gaussian initial data with a large width \(\sigma=0.75\) so
that the data are non-negligible at \(\rho=1\).

\begin{figure}[t]
  \centering
  \includegraphics[width=\textwidth]{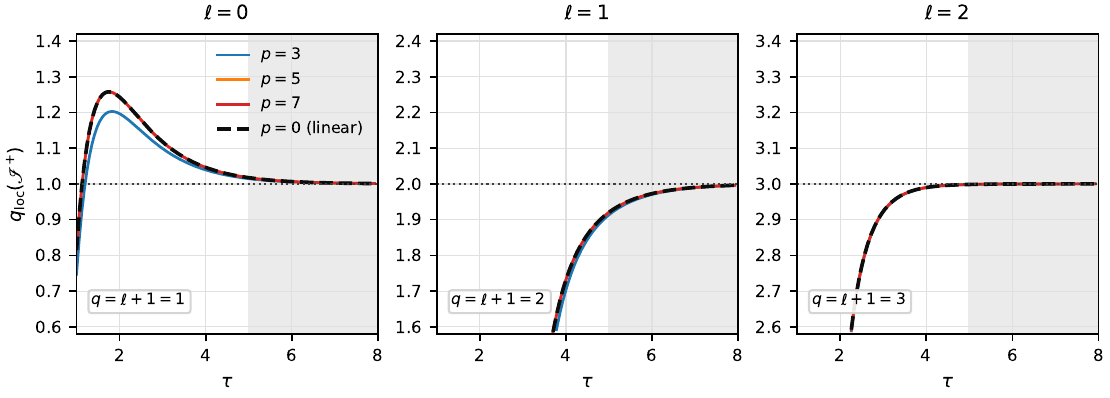}
  \caption{Local decay rates at \(\scri\) for non-compact
  hyperboloidal data with non-negligible support at the boundary.
  Each panel shows the rates for the angular mode \(\ell=0,1,2\) 
  for the linear evolution, labelled \(p=0\), with the semilinear
  evolutions \(p=3,5,7\). The initial data are wide Gaussians with \(A=10\),
  \(\sigma=0.75\), and \(m=0\).  The evolutions use homothetic
  coordinates with \(N_r=64\), \(N_\theta=16\), \(N_\vphi=4\), and
  \(\Delta\tau=0.02\), and run to \(\tau=8.1\). The
  shaded region marks the late-time window \(\tau\in[5,8]\).  The
  dotted horizontal line is \(q=\ell+1\).  The agreement between
  linear and semilinear curves shows that this slow null-infinity
  decay is controlled by the asymptotic support of the data, not by
  the nonlinear power.}
  \label{fig:noncompact-scri-tail-survey}
\end{figure}

Figure~\ref{fig:noncompact-scri-tail-survey} shows that the local rates
for \(p=3,5,7\) coincide with the linear evolution and
approach
\begin{equation}
  q_{\scri}^{\rm nc}(\ell)=\ell+1,
  \qquad \ell=0,1,2 .
\end{equation}
The curve labelled \(p=0\) denotes the linear equation without the semilinear source terms.

One may intuitively expect that there should be no tail for the linear case due to the Huygens'
principle. However, the observed decay is not a tail in the usual sense.  The leading signal at late
times is a linear contribution from the asymptotic data, which decays more slowly than the nonlinear
tails for compactly supported data. As the initial data extend to \(\scri\), in
physical radius they extend to arbitrarily large \(r\) and later retarded times at null infinity. 
Therefore, there is no final outgoing wavefront after which the linear
radiation field must vanish.  The leading signal is instead a linear
asymptotic-data contribution and any nonlinear tail is subleading for this data class.

This behavior is consistent with the distinction between linear and
nonlinear tail mechanisms discussed in perturbative studies of nonlinear
waves \cite{Szpak:2008jv, Bizon:2008ew}.  It is also related to
the Newman--Penrose charge hierarchy, and more broadly to descriptions of
late-time tails in terms of radiation fields at null infinity
\cite{Angelopoulos:2016wcv, Luk:2024ghv}.  The non-vanishing boundary data excite the corresponding
linear asymptotic hierarchy.

\subsection{Higher-mode survey}
\label{sec:higher-mode-survey}

\begin{table}[t]
  \centering    
  \caption{Higher-mode homothetic tail survey at $\scri$ for
  \(p=3,5,7\) and \(\ell=0,1,2\).  The expected rates are
  \(\tilde q_\ell=\max(p-2,\ell+1)\).
  The runs use initial data with the amplitudes $A$ listed in the table,
  \(\sigma=0.2\), \(m=0\); angular resolutions
  \((N_\theta,N_\vphi)=(16,4)\), timestep \(\Delta\tau=0.0025\), and
  \(\tau_{\rm end}=7.1\). The local power index is computed on the
  interval \([\tau_1,\tau_2]_{\rm fit}\).}
  \begin{tabular}{c|c|c|ccc|c}        
    \hline\hline
    $p$ & $A$ & $N_r$ & $\ell=0$ & $\ell=1$ & $\ell=2$ & $[\tau_1,\tau_2]_{\rm fit}$ \\
    \hline
    3 & 40 & 64  & 1.003 & 2.042 & 1.207 & [4.5,7]\\
    3 & 40 & 128 & 1.003 & 2.015 & 3.039 & [4.5,7]\\
    3 & 40 & 256 & 1.003 & 2.015 & 3.019 & [4.5,7]\\
    \hline
    5 & 40 & 64  & 2.871 & 0.447 & -0.147 & [4.5,7]\\
    5 & 40 & 128 & 3.025 & 3.052 & 3.112 & [4.5,7]\\
    5 & 40 & 256 & 3.021 & 3.050 & 3.028 & [4.5,7]\\
    \hline
    7 & 58 & 64  & 5.661 & 4.894 & 4.073 & [3,4.5]\\
    7 & 58 & 128 & 5.343 & 5.376 & 5.502 & [3,4.5]\\
    7 & 58 & 256 & 5.288 & 5.255 & 5.275 & [3,5]\\
    \hline\hline
  \end{tabular}
 \label{tab:higher-mode-survey}
\end{table}

The stationary--homothetic comparisons above demonstrate the main advantage
of the homothetic formulation: once the solution is in the asymptotic
tail regime, the decay exponent is the same at every fixed
compactified radius and
long retarded-time intervals are reached logarithmically in the
evolution time.  We now use homothetic evolutions to test the
expected null-infinity rates for several nonlinear powers and angular
modes. 

Table~\ref{tab:higher-mode-survey} summarizes the measured late-time
local indices at increasing radial resolution.  At \(N_r=256\), the
\(p=3\) rates agree with the expected values to within \(1\%\), and
the \(p=5\) rates to within \(2\%\).  The three \(p=7\) rates lie
between \(5.255\) and \(5.288\), about \(5\)--\(6\%\) above the
predicted uniform rate \(5\).

\subsection{Numerical evidence for a nongeneric cubic monopole cancellation}
\label{sec:non-generic-counterexample}

There is ample evidence supporting Rinne's conjecture for the decay rates of higher multipoles.
However, the conjecture should be understood as a statement about the generic behavior for compactly supported data, not
necessarily the behavior for all smooth data sets. In the following, we construct a numerical
example of a nongeneric cubic monopole tail, in which the leading $e^{-\tau}$ coefficient at null infinity vanishes
and the observed decay rate is close to $e^{-2\tau}$, one exponent faster than the generic tail.
This provides numerical evidence consistent with a nongeneric
codimension-one cancellation of the leading monopole coefficient.

The generic null-infinity prediction for the cubic monopole is $\tilde q_0=1$, but the leading
amplitude
of that tail is a functional of the initial data.  A generic-rate
statement means that this functional is nonzero on an open dense set;
it does not mean that the functional is nonzero for every smooth data
set. For the cubic equation, $p=3$, the expected generic monopole behavior
at $\scri$ is
\begin{equation}\label{eq:non-generic-monopole-expansion}
  a_{00}(\tau;\lambda)
  =
  A_1(\lambda)e^{-\tau}
  +
  A_2(\lambda)e^{-2\tau}
  +
  \O(e^{-3\tau}),
\end{equation}
where $a_{00}$ denotes the spherical-harmonic coefficient of the
rescaled field at $\rho=1$.  If the coefficient $A_1$ vanishes for
special data, the next term in the expansion controls the signal.
For such data the measured exponent is expected to be one exponent 
faster than the generic monopole tail.

We use an axisymmetric one-parameter family of initial
data in homothetic coordinates,
\begin{equation}\label{eq:p3-counterexample-data}
  \psi(0,\rho,\theta,\varphi)=0,
  \qquad
  \partial_\tau\psi(0,\rho,\theta,\varphi)
  =
  A\,\rho^4
  \exp\!\left[-\left(\frac{\rho-\rho_0}{\delta_\rho}\right)^2\right]
  \left(Y_{20}(\theta,\varphi)+\lambda Y_{00}(\theta,\varphi)\right),
\end{equation}
with real normalized spherical harmonics.  In the runs reported below
we set $A=50$, $\rho_0=0.3$, $\delta_\rho=0.08$.
With these parameters, the initial data are numerically negligible near
the origin.
The common factor $\rho^4$ makes both the monopole and quadrupole
pieces regular at the origin.  The
parameter $\lambda$ controls a small monopole mix to otherwise
quadrupolar data. For this
smooth one-parameter family, the leading nonlinear tail coefficient
$A_1$ in \eqref{eq:non-generic-monopole-expansion} is expected to
depend continuously on $\lambda$. If two nearby values of $\lambda$
give opposite signs of $A_1$, then there must be a $\lambda_\ast$ 
for which $A_1(\lambda_\ast)=0$.  
The transverse zero found within this one-parameter family is
consistent with the family intersecting a codimension-one cancellation
surface in the full space of initial data.  The observed monopole
signal is governed by
$A_2 e^{-2\tau}$ until higher-order asymptotic terms or numerical
effects become relevant.  An arbitrarily small perturbation of the
tuned data generically restores a nonzero $A_1$ and the
$e^{-\tau}$ tail.

Numerically, for each $\lambda$ we evolve the cubic equation in
homothetic coordinates and extract the monopole coefficient $a_{00}$
at $\scri$ ($\rho=1$).  On a late-time fitting window
$[\tau_1,\tau_2]$ we estimate the leading coefficient by fitting the signal to
\begin{equation}\label{eq:A1-fit}
  e^\tau a_{00}(\tau;\lambda)
  =
  A_1(\lambda)+B(\lambda)e^{-\tau}.
\end{equation}
This fit is tailored to \eqref{eq:non-generic-monopole-expansion}. The intercept 
is the coefficient of the conjectured generic
$e^{-\tau}$ tail.  We then bisect in $\lambda$ using the sign
of this fitted intercept.  
At resolution
$(N_r,N_\theta,N_\vphi)=(144,10,4)$, with
$\Delta\tau=0.005$, $\tau_{\rm end}=7.1$, and fitting window
$[6.8,7.0]$, the bisection yields
$\lambda_\ast = 0.0007546$.
At this value the fitted leading coefficient is
$A_1=-8.8\times 10^{-15}$.  On the same fitting window the tuned run
gives $q_{\rm fit}=2.004$, while the generic comparison run with
$\lambda=1$ gives $q_{\rm fit}=1.002$.
Additional radial-resolution and timestep experiments confirm that
the cancellation persists, with a comparable nonzero crossing slope
in every tested configuration.  The fitted value of $\lambda_\ast$
has a small resolution dependence, as expected for a numerically
determined cancellation threshold.

\begin{figure}[t]
  \centering
  \includegraphics[width=0.48\textwidth]{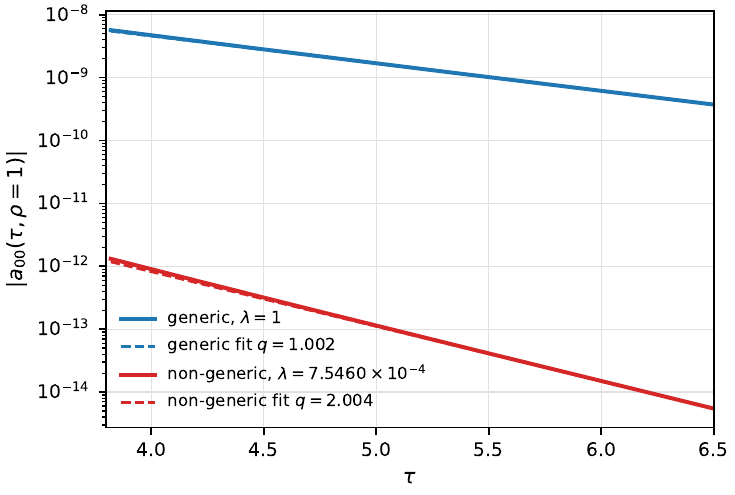}
  \hfill
  \includegraphics[width=0.48\textwidth]{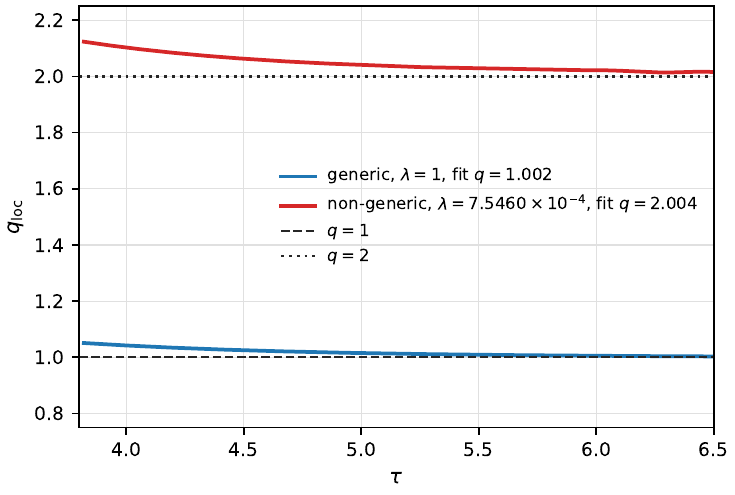}
  \caption{Generic and nongeneric cubic monopole tails at $\scri$.
  Left: monopole signal $|a_{00}(\tau,\rho=1)|$ for generic and
  tuned data, with dashed least-squares exponential fits.  The plotted curves are shown only up to
  $\tau=6.5$.  Right: local decay rate
  $q_{\rm loc}=-\partial_\tau\ln |a_{00}|$ for the same data, with
  reference lines at $q=1$ and $q=2$.  The generic data
  ($\lambda=1$) give $q_{\rm fit}=1.002$, while the tuned data
  ($\lambda=0.0007546$) give $q_{\rm fit}=2.004$.}
  \label{fig:p3-counterexample-comparison}
\end{figure}

Figure~\ref{fig:p3-counterexample-comparison} shows a comparison 
between the generic and the nongeneric cases. The generic case, represented
by $\lambda=1$, has a strong monopole signal and a local decay rate
near $q=1$.  The tuned case has a much smaller monopole signal, and
the local decay rate is close to $q=2$ throughout the fitting window.
This tuning provides numerical evidence consistent with a
nongeneric codimension-one cancellation of the leading monopole
coefficient.


\section{Conclusions}
\label{sec:conclusions}

We presented a homothetic, hyperboloidal formulation for the computation of late-time tails in semilinear wave equations. The central result is that this formulation avoids the gradient steepening of stationary hyperboloidal compactifications. In a stationary scri-fixing foliation, a single compactified slice must represent both the faster decay measured at finite radius and the slower radiative decay at \(\scri\). As time increases, these different power laws produce an increasingly narrow transition near the compactified boundary and therefore an increasingly more difficult spatial-resolution problem. We showed that this behavior is avoidable, as it results from representing a self-similar asymptotic solution in coordinates adapted to time translations.

The homothetic coordinates
\cite{Donninger:2013sba,Nutzi:2023uly,Nutzi:2025kqc} are instead
adapted to the scaling structure of the tail.  Suppressing the
$m$-index, an asymptotic radiative mode takes the form $\chi_\ell(\tau,\rho)
  \sim e^{-\widetilde q_\ell\tau}P_\ell(\rho)$
at every fixed $0<\rho\leq1$, with a time-independent radial profile
and the same decay exponent at every fixed compactified radius.  At a
fixed physical radius $r>0$, letting $t\to\infty$ implies $\rho=r/t\to0$, $u\sim t$, $\tau\to\infty$.
The origin-regularity factor then recovers the faster finite-radius
decay, rather than requiring a change of exponent across fixed grid
points.  Therefore, the normalized tail profile does not develop the
progressively sharpening boundary layer characteristic of stationary
hyperboloidal evolution.  This removal of the late-time gradient
problem is the main computational advantage of the homothetic
formulation for the study of late-time tails.

We presented numerical evidence for this mechanism using a
$3+1$-dimensional pseudospectral implementation.  The stationary
evolutions develop increasingly steep profiles near $\scri$, whereas
the homothetic profiles approach a smooth shape with a uniform decay rate throughout the compactified domain.  Since the
retarded time at $\scri$ satisfies $u=Re^\tau$, uniform steps in
homothetic time cover exponentially large intervals.
The spatial representation remains
resolved because the asymptotic profile approaches a fixed smooth
shape instead of developing sharp gradients.

We used this framework to investigate the decay-rate conjecture proposed by Rinne for higher multipoles of semilinear waves. For compactly supported or sufficiently rapidly decaying data, our results provide further evidence for the conjectured generic rates. However, the numerical experiments show that the relevant data class must be specified. When the hyperboloidal initial data are non-negligible at \(\scri\), the leading signal is controlled by the linear asymptotic data rather than by the nonlinear compact-data tail. For the examples studied here, we get $q_{\scri}^{\rm nc}(\ell)=\ell+1$ for $\ell=0,1,2$,
independently of the nonlinear powers \(p=3,5,7\).

We also constructed a one-parameter family of solutions for the cubic equation in which the fitted
coefficient of the generic $e^{-\tau}$ monopole tail changes sign.
Tuning to the zero of this coefficient suppresses the generic
contribution and exposes a resolved $e^{-2\tau}$ regime. Nearby
untuned data recover the $e^{-\tau}$ behavior.  This demonstrates
that the conjectured rates are generic, not universal, and provides
numerical evidence consistent with a codimension-one cancellation of
the leading monopole coefficient.

The homothetic hyperboloidal construction has two natural extensions. For Schwarzschild and Kerr spacetimes, the black-hole mass introduces an intrinsic scale, so a globally homothetic foliation is not expected. A promising strategy is instead to use stationary horizon-penetrating coordinates in the strong-field interior and match them to an asymptotically homothetic compactification in the exterior, in the spirit of earlier matched hyperboloidal constructions \cite{Zenginoglu:2007jw,Zenginoglu:2009hd}. Such a formulation could retain regular access to both the event horizon and \(\scri\) while avoiding the asymptotic gradient steepening addressed in this work.

A motivation for this study is the global numerical
evolution of the full nonlinear vacuum Einstein equations for small
asymptotically flat data with scri-fixing.  The next stage of this
program is a $3+1$-dimensional
evolution on the full spatial ball without symmetry assumptions,
based on N\"utzi's regular formulation \cite{Nutzi:2025kqc}.  Such a
calculation requires a regular treatment of the conformal boundary, a
discretization of the full ball, and coordinates that remain resolved over
very long retarded times. We established two of these numerical
ingredients in the case of the semilinear wave equation: a regular full-ball
spectral discretization and a scaling-adapted foliation that avoids
the late-time steepening of stationary hyperboloidal
coordinates.

\section*{Code and data availability}

The code and data used for the numerical experiments will be made
available on GitHub. 

\begin{acknowledgments}
  AZ is supported by the National Science Foundation under Grant
  No. 2309084.
  SB acknowledges funding from the EU Horizon under ERC Consolidator
  Grant, no. InspiReM-101043372.
\end{acknowledgments}


\appendix


\section{Numerical method}\label{app:numerics}

The semilinear wave equations in our work are solved with spectral
methods based on the Dedalus~\footnote{\url{https://dedalus-project.org/}}
library~\cite{Burns:2020}. The latter provides, among other features,
a ``ball'' basis to solve PDEs on
the full sphere using (tensor) spherical harmonics in the angular
directions and a scaled Jacobi polynomial basis in the radial
direction~\cite{Lecoanet:2019,Vasil:2018rsw}.
Nonlinear terms are calculated by transforming from the coefficients
to physical grid. Time integration is performed with the
implicit-explicit 3rd-order 4-stage DIRK+ERK (RK443) scheme. The 
implicit step of the scheme is applied to the linear part of the
spatial operator, while the nonlinear terms are handled
efficiently with the explicit step. 
This provides temporal stability for linearly stiff equations without
requiring iterative algorithms for integrating the nonlinear terms.

For the numerical evolution we work with the first-order variables
\[
U=\psi,\qquad
\Pi=\partial_\tau \psi,\qquad
W=\mathcal D_\rho\psi,\qquad
\mathcal D_\rho:=\rho\,\partial_\rho,
\]
and solve the system on the Dedalus ball using a spherical spectral
discretization.

\subsection{Regularity at the origin}
The second-order equations \eqref{eq:poincare_equation} and
\eqref{eq:homothetic_wave} have coefficients of the form
\[
\frac{1}{\rho}\,(\cdots),\qquad \frac{1}{\rho^2}\,(\cdots)
.
\]
In a direct implementation, regularity is obtained through
cancellations between radial and angular terms. To make the origin
regularity explicit in the numerical formulation, we rewrite the
equation to make the Laplacian manifest, 
\[
\Delta U
=
U_{\rho\rho}+\frac{2}{\rho}U_\rho+\frac{1}{\rho^2}\Delta_{\mathbb S^2}U.
\]
The implemented systems use \(\Delta U\) as a single Dedalus ball
operator, rather than evaluating the radial and angular singular terms
separately.  The auxiliary relation \(W=\mathcal D_\rho U\) is imposed
as a first-order equation, where
\(\mathcal D_\rho=\rho\partial_\rho\).  This gives a system whose
coefficients are regular on the closed ball and whose origin
regularity is handled by the ball basis.

For the homothetic formulation, the implemented equations have the
schematic form
\begin{align}
  \partial_\tau U &= \Pi,\\
  W &= \mathcal D_\rho U,\\
  \partial_\tau \Pi
  &=
  \Delta U-\mathcal D_\rho W-3W
  -2\mathcal D_\rho\Pi-3\Pi-2U
  +S_{\rm hom}(\tau,\rho,U).
\end{align}
Here
\[
  S_{\rm hom}(\tau,\rho,U)
  =
  \pm e^{-(p-3)\tau}
  \left(\frac{1-\rho^2}{2}\right)^{p-3}U^p,
\]
with the sign chosen according to the focusing or defocusing
nonlinearity.  For the stationary hyperboloidal formulation we use the
same variables and write the equation in the corresponding regular
form
\begin{align}
  \partial_\tau U &= \Pi,\\
  W &= \mathcal D_\rho U,\\
  \partial_\tau \Pi
  &=
  D(\rho)\Delta U-\mathcal D_\rho W
  +C_\Pi(\rho)\Pi+C_W(\rho)W+C_U(\rho)U
  +S_{\rm stat}(\rho,U),
\end{align}
where \(D,C_\Pi,C_W,C_U\) are smooth functions of \(\rho\) on
\([0,1]\), and
\[
  S_{\rm stat}(\rho,U)
  =
  \pm \frac{(1+\rho^2)^2}{4}
  \left(\frac{1-\rho^2}{2}\right)^{p-3}U^p .
\]
The coefficient functions are generated in the
solver from the formulas in Secs.~\ref{sec:stationary} and
\ref{sec:homothetic}.  The
singular \(1/\rho\) and \(1/\rho^2\) factors in the coordinate
expressions never appear as separate grid operations.  Instead, the
regular ball Laplacian and the operator \(\mathcal D_\rho\) encode the
origin behavior in a form compatible with the spectral basis.


\begin{thebibliography}{55}%
\makeatletter
\providecommand \@ifxundefined [1]{%
 \@ifx{#1\undefined}
}%
\providecommand \@ifnum [1]{%
 \ifnum #1\expandafter \@firstoftwo
 \else \expandafter \@secondoftwo
 \fi
}%
\providecommand \@ifx [1]{%
 \ifx #1\expandafter \@firstoftwo
 \else \expandafter \@secondoftwo
 \fi
}%
\providecommand \natexlab [1]{#1}%
\providecommand \enquote  [1]{``#1''}%
\providecommand \bibnamefont  [1]{#1}%
\providecommand \bibfnamefont [1]{#1}%
\providecommand \citenamefont [1]{#1}%
\providecommand \href@noop [0]{\@secondoftwo}%
\providecommand \href [0]{\begingroup \@sanitize@url \@href}%
\providecommand \@href[1]{\@@startlink{#1}\@@href}%
\providecommand \@@href[1]{\endgroup#1\@@endlink}%
\providecommand \@sanitize@url [0]{\catcode `\\12\catcode `\$12\catcode
  `\&12\catcode `\#12\catcode `\^12\catcode `\_12\catcode `\%12\relax}%
\providecommand \@@startlink[1]{}%
\providecommand \@@endlink[0]{}%
\providecommand \url  [0]{\begingroup\@sanitize@url \@url }%
\providecommand \@url [1]{\endgroup\@href {#1}{\urlprefix }}%
\providecommand \urlprefix  [0]{URL }%
\providecommand \Eprint [0]{\href }%
\providecommand \doibase [0]{http://dx.doi.org/}%
\providecommand \selectlanguage [0]{\@gobble}%
\providecommand \bibinfo  [0]{\@secondoftwo}%
\providecommand \bibfield  [0]{\@secondoftwo}%
\providecommand \translation [1]{[#1]}%
\providecommand \BibitemOpen [0]{}%
\providecommand \bibitemStop [0]{}%
\providecommand \bibitemNoStop [0]{.\EOS\space}%
\providecommand \EOS [0]{\spacefactor3000\relax}%
\providecommand \BibitemShut  [1]{\csname bibitem#1\endcsname}%
\let\auto@bib@innerbib\@empty
\bibitem [{\citenamefont {Price}(1972)}]{Price:1972pw}%
  \BibitemOpen
  \bibfield  {author} {\bibinfo {author} {\bibfnamefont {R.~H.}\ \bibnamefont
  {Price}},\ }\href {\doibase 10.1103/PhysRevD.5.2439} {\bibfield  {journal}
  {\bibinfo  {journal} {Phys. Rev.}\ }\textbf {\bibinfo {volume} {D5}},\
  \bibinfo {pages} {2439} (\bibinfo {year} {1972})}\BibitemShut {NoStop}%
\bibitem [{\citenamefont {Bizon}\ \emph {et~al.}(2008)\citenamefont {Bizon},
  \citenamefont {Chmaj},\ and\ \citenamefont {Rostworowski}}]{Bizon:2008ew}%
  \BibitemOpen
  \bibfield  {author} {\bibinfo {author} {\bibfnamefont {P.}~\bibnamefont
  {Bizon}}, \bibinfo {author} {\bibfnamefont {T.}~\bibnamefont {Chmaj}}, \ and\
  \bibinfo {author} {\bibfnamefont {A.}~\bibnamefont {Rostworowski}},\ }\href
  {\doibase 10.1103/PhysRevD.78.024044} {\bibfield  {journal} {\bibinfo
  {journal} {Phys. Rev. D}\ }\textbf {\bibinfo {volume} {78}},\ \bibinfo
  {pages} {024044} (\bibinfo {year} {2008})},\ \Eprint
  {http://arxiv.org/abs/0804.0903} {arXiv:0804.0903 [math-ph]} \BibitemShut
  {NoStop}%
\bibitem [{\citenamefont {Szpak}\ \emph {et~al.}(2009)\citenamefont {Szpak},
  \citenamefont {Bizon}, \citenamefont {Chmaj},\ and\ \citenamefont
  {Rostworowski}}]{Szpak:2008jv}%
  \BibitemOpen
  \bibfield  {author} {\bibinfo {author} {\bibfnamefont {N.}~\bibnamefont
  {Szpak}}, \bibinfo {author} {\bibfnamefont {P.}~\bibnamefont {Bizon}},
  \bibinfo {author} {\bibfnamefont {T.}~\bibnamefont {Chmaj}}, \ and\ \bibinfo
  {author} {\bibfnamefont {A.}~\bibnamefont {Rostworowski}},\ }\href {\doibase
  10.1142/S0219891609001782} {\bibfield  {journal} {\bibinfo  {journal} {J.
  Hyperbol. Diff. Equat.}\ }\textbf {\bibinfo {volume} {6}},\ \bibinfo {pages}
  {107} (\bibinfo {year} {2009})},\ \Eprint {http://arxiv.org/abs/0712.0493}
  {arXiv:0712.0493 [math-ph]} \BibitemShut {NoStop}%
\bibitem [{\citenamefont {Harms}\ \emph {et~al.}(2013)\citenamefont {Harms},
  \citenamefont {Bernuzzi},\ and\ \citenamefont {Br{\"u}gmann}}]{Harms:2013ib}%
  \BibitemOpen
  \bibfield  {author} {\bibinfo {author} {\bibfnamefont {E.}~\bibnamefont
  {Harms}}, \bibinfo {author} {\bibfnamefont {S.}~\bibnamefont {Bernuzzi}}, \
  and\ \bibinfo {author} {\bibfnamefont {B.}~\bibnamefont {Br{\"u}gmann}},\
  }\href {\doibase 10.1088/0264-9381/30/11/115013} {\bibfield  {journal}
  {\bibinfo  {journal} {Class.Quant.Grav.}\ }\textbf {\bibinfo {volume} {30}},\
  \bibinfo {pages} {115013} (\bibinfo {year} {2013})},\ \Eprint
  {http://arxiv.org/abs/1301.1591} {arXiv:1301.1591 [gr-qc]} \BibitemShut
  {NoStop}%
\bibitem [{\citenamefont {Albanesi}\ \emph {et~al.}(2023)\citenamefont
  {Albanesi}, \citenamefont {Bernuzzi}, \citenamefont {Damour}, \citenamefont
  {Nagar},\ and\ \citenamefont {Placidi}}]{Albanesi:2023bgi}%
  \BibitemOpen
  \bibfield  {author} {\bibinfo {author} {\bibfnamefont {S.}~\bibnamefont
  {Albanesi}}, \bibinfo {author} {\bibfnamefont {S.}~\bibnamefont {Bernuzzi}},
  \bibinfo {author} {\bibfnamefont {T.}~\bibnamefont {Damour}}, \bibinfo
  {author} {\bibfnamefont {A.}~\bibnamefont {Nagar}}, \ and\ \bibinfo {author}
  {\bibfnamefont {A.}~\bibnamefont {Placidi}},\ }\href {\doibase
  10.1103/PhysRevD.108.084037} {\bibfield  {journal} {\bibinfo  {journal}
  {Phys. Rev. D}\ }\textbf {\bibinfo {volume} {108}},\ \bibinfo {pages}
  {084037} (\bibinfo {year} {2023})},\ \Eprint
  {http://arxiv.org/abs/2305.19336} {arXiv:2305.19336 [gr-qc]} \BibitemShut
  {NoStop}%
\bibitem [{\citenamefont {De~Amicis}\ \emph {et~al.}(2024)\citenamefont
  {De~Amicis}, \citenamefont {Albanesi},\ and\ \citenamefont
  {Carullo}}]{DeAmicis:2024not}%
  \BibitemOpen
  \bibfield  {author} {\bibinfo {author} {\bibfnamefont {M.}~\bibnamefont
  {De~Amicis}}, \bibinfo {author} {\bibfnamefont {S.}~\bibnamefont {Albanesi}},
  \ and\ \bibinfo {author} {\bibfnamefont {G.}~\bibnamefont {Carullo}},\ }\href
  {\doibase 10.1103/PhysRevD.110.104005} {\bibfield  {journal} {\bibinfo
  {journal} {Phys. Rev. D}\ }\textbf {\bibinfo {volume} {110}},\ \bibinfo
  {pages} {104005} (\bibinfo {year} {2024})},\ \Eprint
  {http://arxiv.org/abs/2406.17018} {arXiv:2406.17018 [gr-qc]} \BibitemShut
  {NoStop}%
\bibitem [{\citenamefont {Islam}\ \emph {et~al.}(2025)\citenamefont {Islam},
  \citenamefont {Faggioli}, \citenamefont {Khanna}, \citenamefont {Field},
  \citenamefont {van~de Meent},\ and\ \citenamefont
  {Buonanno}}]{Islam:2024vro}%
  \BibitemOpen
  \bibfield  {author} {\bibinfo {author} {\bibfnamefont {T.}~\bibnamefont
  {Islam}}, \bibinfo {author} {\bibfnamefont {G.}~\bibnamefont {Faggioli}},
  \bibinfo {author} {\bibfnamefont {G.}~\bibnamefont {Khanna}}, \bibinfo
  {author} {\bibfnamefont {S.~E.}\ \bibnamefont {Field}}, \bibinfo {author}
  {\bibfnamefont {M.}~\bibnamefont {van~de Meent}}, \ and\ \bibinfo {author}
  {\bibfnamefont {A.}~\bibnamefont {Buonanno}},\ }\href {\doibase
  10.1103/191t-5svc} {\bibfield  {journal} {\bibinfo  {journal} {Phys. Rev. D}\
  }\textbf {\bibinfo {volume} {112}},\ \bibinfo {pages} {024061} (\bibinfo
  {year} {2025})},\ \Eprint {http://arxiv.org/abs/2407.04682} {arXiv:2407.04682
  [gr-qc]} \BibitemShut {NoStop}%
\bibitem [{\citenamefont {Islam}\ \emph {et~al.}(2026)\citenamefont {Islam},
  \citenamefont {Faggioli},\ and\ \citenamefont {Khanna}}]{Islam:2025wci}%
  \BibitemOpen
  \bibfield  {author} {\bibinfo {author} {\bibfnamefont {T.}~\bibnamefont
  {Islam}}, \bibinfo {author} {\bibfnamefont {G.}~\bibnamefont {Faggioli}}, \
  and\ \bibinfo {author} {\bibfnamefont {G.}~\bibnamefont {Khanna}},\ }\href
  {\doibase 10.1103/5pfc-rnl4} {\bibfield  {journal} {\bibinfo  {journal}
  {Phys. Rev. D}\ }\textbf {\bibinfo {volume} {113}},\ \bibinfo {pages}
  {124025} (\bibinfo {year} {2026})},\ \Eprint
  {http://arxiv.org/abs/2511.21898} {arXiv:2511.21898 [gr-qc]} \BibitemShut
  {NoStop}%
\bibitem [{\citenamefont {Alnasheet}\ \emph {et~al.}(2025)\citenamefont
  {Alnasheet}, \citenamefont {Cardoso}, \citenamefont {Duque},\ and\
  \citenamefont {Panosso~Macedo}}]{Alnasheet:2025mtr}%
  \BibitemOpen
  \bibfield  {author} {\bibinfo {author} {\bibfnamefont {Q.}~\bibnamefont
  {Alnasheet}}, \bibinfo {author} {\bibfnamefont {V.}~\bibnamefont {Cardoso}},
  \bibinfo {author} {\bibfnamefont {F.}~\bibnamefont {Duque}}, \ and\ \bibinfo
  {author} {\bibfnamefont {R.}~\bibnamefont {Panosso~Macedo}},\ }\href
  {\doibase 10.1103/yyv5-3y1c} {\bibfield  {journal} {\bibinfo  {journal}
  {Phys. Rev. D}\ }\textbf {\bibinfo {volume} {112}},\ \bibinfo {pages}
  {044066} (\bibinfo {year} {2025})},\ \Eprint
  {http://arxiv.org/abs/2508.20238} {arXiv:2508.20238 [gr-qc]} \BibitemShut
  {NoStop}%
\bibitem [{\citenamefont {Vega}\ \emph {et~al.}(2026)\citenamefont {Vega},
  \citenamefont {Svyatkovskyy~Kholyavka}, \citenamefont {Datta},\ and\
  \citenamefont {Forteza}}]{Vega:2026lfs}%
  \BibitemOpen
  \bibfield  {author} {\bibinfo {author} {\bibfnamefont {J.~A.~L.}\
  \bibnamefont {Vega}}, \bibinfo {author} {\bibfnamefont {A.}~\bibnamefont
  {Svyatkovskyy~Kholyavka}}, \bibinfo {author} {\bibfnamefont {S.}~\bibnamefont
  {Datta}}, \ and\ \bibinfo {author} {\bibfnamefont {X.~J.}\ \bibnamefont
  {Forteza}},\ }\href@noop {} {\  (\bibinfo {year} {2026})},\ \Eprint
  {http://arxiv.org/abs/2606.02146} {arXiv:2606.02146 [gr-qc]} \BibitemShut
  {NoStop}%
\bibitem [{\citenamefont {De~Amicis}\ \emph {et~al.}(2025)\citenamefont
  {De~Amicis} \emph {et~al.}}]{DeAmicis:2024eoy}%
  \BibitemOpen
  \bibfield  {author} {\bibinfo {author} {\bibfnamefont {M.}~\bibnamefont
  {De~Amicis}} \emph {et~al.},\ }\href {\doibase 10.1103/2brx-xnyr} {\bibfield
  {journal} {\bibinfo  {journal} {Phys. Rev. Lett.}\ }\textbf {\bibinfo
  {volume} {135}},\ \bibinfo {pages} {171401} (\bibinfo {year} {2025})},\
  \Eprint {http://arxiv.org/abs/2412.06887} {arXiv:2412.06887 [gr-qc]}
  \BibitemShut {NoStop}%
\bibitem [{\citenamefont {Ling}\ \emph {et~al.}(2025)\citenamefont {Ling},
  \citenamefont {Shah},\ and\ \citenamefont {Wong}}]{Ling:2025wfv}%
  \BibitemOpen
  \bibfield  {author} {\bibinfo {author} {\bibfnamefont {S.}~\bibnamefont
  {Ling}}, \bibinfo {author} {\bibfnamefont {S.}~\bibnamefont {Shah}}, \ and\
  \bibinfo {author} {\bibfnamefont {S.~S.~C.}\ \bibnamefont {Wong}},\ }\href
  {\doibase 10.1103/22lc-62gj} {\bibfield  {journal} {\bibinfo  {journal}
  {Phys. Rev. D}\ }\textbf {\bibinfo {volume} {112}},\ \bibinfo {pages}
  {024008} (\bibinfo {year} {2025})},\ \Eprint
  {http://arxiv.org/abs/2503.19967} {arXiv:2503.19967 [gr-qc]} \BibitemShut
  {NoStop}%
\bibitem [{\citenamefont {Ling}\ and\ \citenamefont
  {Wong}(2026)}]{Ling:2026ynd}%
  \BibitemOpen
  \bibfield  {author} {\bibinfo {author} {\bibfnamefont {S.}~\bibnamefont
  {Ling}}\ and\ \bibinfo {author} {\bibfnamefont {S.~S.~C.}\ \bibnamefont
  {Wong}},\ }\href@noop {} {\  (\bibinfo {year} {2026})},\ \Eprint
  {http://arxiv.org/abs/2603.20379} {arXiv:2603.20379 [gr-qc]} \BibitemShut
  {NoStop}%
\bibitem [{\citenamefont {Zengino{\u
  g}lu}(2008{\natexlab{a}})}]{Zenginoglu:2008wc}%
  \BibitemOpen
  \bibfield  {author} {\bibinfo {author} {\bibfnamefont {A.}~\bibnamefont
  {Zengino{\u g}lu}},\ }\href {\doibase 10.1088/0264-9381/25/17/175013}
  {\bibfield  {journal} {\bibinfo  {journal} {Class. Quant. Grav.}\ }\textbf
  {\bibinfo {volume} {25}},\ \bibinfo {pages} {175013} (\bibinfo {year}
  {2008}{\natexlab{a}})},\ \Eprint {http://arxiv.org/abs/0803.2018}
  {arXiv:0803.2018 [gr-qc]} \BibitemShut {NoStop}%
\bibitem [{\citenamefont {Zengino{\u
  g}lu}(2008{\natexlab{b}})}]{Zenginoglu:2007jw}%
  \BibitemOpen
  \bibfield  {author} {\bibinfo {author} {\bibfnamefont {A.}~\bibnamefont
  {Zengino{\u g}lu}},\ }\href {\doibase 10.1088/0264-9381/25/14/145002}
  {\bibfield  {journal} {\bibinfo  {journal} {Class. Quant. Grav.}\ }\textbf
  {\bibinfo {volume} {25}},\ \bibinfo {pages} {145002} (\bibinfo {year}
  {2008}{\natexlab{b}})},\ \Eprint {http://arxiv.org/abs/0712.4333}
  {arXiv:0712.4333 [gr-qc]} \BibitemShut {NoStop}%
\bibitem [{\citenamefont {Zengino{\u g}lu}\ \emph {et~al.}(2009)\citenamefont
  {Zengino{\u g}lu}, \citenamefont {Nunez},\ and\ \citenamefont
  {Husa}}]{Zenginoglu:2008uc}%
  \BibitemOpen
  \bibfield  {author} {\bibinfo {author} {\bibfnamefont {A.}~\bibnamefont
  {Zengino{\u g}lu}}, \bibinfo {author} {\bibfnamefont {D.}~\bibnamefont
  {Nunez}}, \ and\ \bibinfo {author} {\bibfnamefont {S.}~\bibnamefont {Husa}},\
  }\href {\doibase 10.1088/0264-9381/26/3/035009} {\bibfield  {journal}
  {\bibinfo  {journal} {Class. Quant. Grav.}\ }\textbf {\bibinfo {volume}
  {26}},\ \bibinfo {pages} {035009} (\bibinfo {year} {2009})},\ \Eprint
  {http://arxiv.org/abs/0810.1929} {arXiv:0810.1929 [gr-qc]} \BibitemShut
  {NoStop}%
\bibitem [{\citenamefont {Zenginoglu}(2010)}]{Zenginoglu:2009ey}%
  \BibitemOpen
  \bibfield  {author} {\bibinfo {author} {\bibfnamefont {A.}~\bibnamefont
  {Zenginoglu}},\ }\href {\doibase 10.1088/0264-9381/27/4/045015} {\bibfield
  {journal} {\bibinfo  {journal} {Class. Quant. Grav.}\ }\textbf {\bibinfo
  {volume} {27}},\ \bibinfo {pages} {045015} (\bibinfo {year} {2010})},\
  \Eprint {http://arxiv.org/abs/0911.2450} {arXiv:0911.2450 [gr-qc]}
  \BibitemShut {NoStop}%
\bibitem [{\citenamefont {Jasiulek}(2012)}]{Jasiulek:2011ce}%
  \BibitemOpen
  \bibfield  {author} {\bibinfo {author} {\bibfnamefont {M.}~\bibnamefont
  {Jasiulek}},\ }\href {\doibase 10.1088/0264-9381/29/1/015008} {\bibfield
  {journal} {\bibinfo  {journal} {Class.Quant.Grav.}\ }\textbf {\bibinfo
  {volume} {29}},\ \bibinfo {pages} {015008} (\bibinfo {year} {2012})},\
  \Eprint {http://arxiv.org/abs/1109.2513} {arXiv:1109.2513 [gr-qc]}
  \BibitemShut {NoStop}%
\bibitem [{\citenamefont {R{\'a}cz}\ and\ \citenamefont
  {T{\'o}th}(2011)}]{Racz:2011qu}%
  \BibitemOpen
  \bibfield  {author} {\bibinfo {author} {\bibfnamefont {I.}~\bibnamefont
  {R{\'a}cz}}\ and\ \bibinfo {author} {\bibfnamefont {G.~Z.}\ \bibnamefont
  {T{\'o}th}},\ }\href {\doibase 10.1088/0264-9381/28/19/195003} {\bibfield
  {journal} {\bibinfo  {journal} {Class.Quant.Grav.}\ }\textbf {\bibinfo
  {volume} {28}},\ \bibinfo {pages} {195003} (\bibinfo {year} {2011})},\
  \Eprint {http://arxiv.org/abs/1104.4199} {arXiv:1104.4199 [gr-qc]}
  \BibitemShut {NoStop}%
\bibitem [{\citenamefont {Bernuzzi}\ \emph {et~al.}(2012)\citenamefont
  {Bernuzzi}, \citenamefont {Nagar},\ and\ \citenamefont
  {Zenginoglu}}]{Bernuzzi:2012ku}%
  \BibitemOpen
  \bibfield  {author} {\bibinfo {author} {\bibfnamefont {S.}~\bibnamefont
  {Bernuzzi}}, \bibinfo {author} {\bibfnamefont {A.}~\bibnamefont {Nagar}}, \
  and\ \bibinfo {author} {\bibfnamefont {A.}~\bibnamefont {Zenginoglu}},\
  }\href {\doibase 10.1103/PhysRevD.86.104038} {\bibfield  {journal} {\bibinfo
  {journal} {Phys.Rev.}\ }\textbf {\bibinfo {volume} {D86}},\ \bibinfo {pages}
  {104038} (\bibinfo {year} {2012})},\ \Eprint {http://arxiv.org/abs/1207.0769}
  {arXiv:1207.0769 [gr-qc]} \BibitemShut {NoStop}%
\bibitem [{\citenamefont {Zenginoğlu}\ \emph {et~al.}(2014)\citenamefont
  {Zenginoğlu}, \citenamefont {Khanna},\ and\ \citenamefont
  {Burko}}]{Zenginoglu:2012us}%
  \BibitemOpen
  \bibfield  {author} {\bibinfo {author} {\bibfnamefont {A.}~\bibnamefont
  {Zenginoğlu}}, \bibinfo {author} {\bibfnamefont {G.}~\bibnamefont {Khanna}},
  \ and\ \bibinfo {author} {\bibfnamefont {L.~M.}\ \bibnamefont {Burko}},\
  }\href {\doibase 10.1007/s10714-014-1672-8} {\bibfield  {journal} {\bibinfo
  {journal} {Gen.Rel.Grav.}\ }\textbf {\bibinfo {volume} {46}},\ \bibinfo
  {pages} {1672} (\bibinfo {year} {2014})},\ \Eprint
  {http://arxiv.org/abs/1208.5839} {arXiv:1208.5839 [gr-qc]} \BibitemShut
  {NoStop}%
\bibitem [{\citenamefont {Harms}\ \emph {et~al.}(2014)\citenamefont {Harms},
  \citenamefont {Bernuzzi}, \citenamefont {Nagar},\ and\ \citenamefont
  {Zenginoglu}}]{Harms:2014dqa}%
  \BibitemOpen
  \bibfield  {author} {\bibinfo {author} {\bibfnamefont {E.}~\bibnamefont
  {Harms}}, \bibinfo {author} {\bibfnamefont {S.}~\bibnamefont {Bernuzzi}},
  \bibinfo {author} {\bibfnamefont {A.}~\bibnamefont {Nagar}}, \ and\ \bibinfo
  {author} {\bibfnamefont {A.}~\bibnamefont {Zenginoglu}},\ }\href {\doibase
  10.1088/0264-9381/31/24/245004} {\bibfield  {journal} {\bibinfo  {journal}
  {Class.Quant.Grav.}\ }\textbf {\bibinfo {volume} {31}},\ \bibinfo {pages}
  {245004} (\bibinfo {year} {2014})},\ \Eprint {http://arxiv.org/abs/1406.5983}
  {arXiv:1406.5983 [gr-qc]} \BibitemShut {NoStop}%
\bibitem [{\citenamefont {Csuk{\'a}s}\ \emph {et~al.}(2019)\citenamefont
  {Csuk{\'a}s}, \citenamefont {R{\'a}cz},\ and\ \citenamefont
  {T{\'o}th}}]{Csukas:2019kcb}%
  \BibitemOpen
  \bibfield  {author} {\bibinfo {author} {\bibfnamefont {K.}~\bibnamefont
  {Csuk{\'a}s}}, \bibinfo {author} {\bibfnamefont {I.}~\bibnamefont
  {R{\'a}cz}}, \ and\ \bibinfo {author} {\bibfnamefont {G.~Z.}\ \bibnamefont
  {T{\'o}th}},\ }\href {\doibase 10.1103/PhysRevD.100.104025} {\bibfield
  {journal} {\bibinfo  {journal} {Phys. Rev. D}\ }\textbf {\bibinfo {volume}
  {100}},\ \bibinfo {pages} {104025} (\bibinfo {year} {2019})},\ \Eprint
  {http://arxiv.org/abs/1905.09082} {arXiv:1905.09082 [gr-qc]} \BibitemShut
  {NoStop}%
\bibitem [{\citenamefont {Zengino{\u g}lu}(2011)}]{Zenginoglu:2010cq}%
  \BibitemOpen
  \bibfield  {author} {\bibinfo {author} {\bibfnamefont {A.}~\bibnamefont
  {Zengino{\u g}lu}},\ }\href {\doibase 10.1016/j.jcp.2010.12.016} {\bibfield
  {journal} {\bibinfo  {journal} {J.Comput.Phys.}\ }\textbf {\bibinfo {volume}
  {230}},\ \bibinfo {pages} {2286} (\bibinfo {year} {2011})},\ \Eprint
  {http://arxiv.org/abs/1008.3809} {arXiv:1008.3809 [math.NA]} \BibitemShut
  {NoStop}%
\bibitem [{\citenamefont {Zengino{\u g}lu}\ and\ \citenamefont
  {Kidder}(2010)}]{Zenginoglu:2010zm}%
  \BibitemOpen
  \bibfield  {author} {\bibinfo {author} {\bibfnamefont {A.}~\bibnamefont
  {Zengino{\u g}lu}}\ and\ \bibinfo {author} {\bibfnamefont {L.~E.}\
  \bibnamefont {Kidder}},\ }\href {\doibase 10.1103/PhysRevD.81.124010}
  {\bibfield  {journal} {\bibinfo  {journal} {Phys. Rev.}\ }\textbf {\bibinfo
  {volume} {D81}},\ \bibinfo {pages} {124010} (\bibinfo {year} {2010})},\
  \Eprint {http://arxiv.org/abs/1004.0760} {arXiv:1004.0760 [gr-qc]}
  \BibitemShut {NoStop}%
\bibitem [{\citenamefont {Hilditch}\ \emph {et~al.}(2018)\citenamefont
  {Hilditch}, \citenamefont {Harms}, \citenamefont {Bugner}, \citenamefont
  {R\"uter},\ and\ \citenamefont {Br\"ugmann}}]{Hilditch:2016xzh}%
  \BibitemOpen
  \bibfield  {author} {\bibinfo {author} {\bibfnamefont {D.}~\bibnamefont
  {Hilditch}}, \bibinfo {author} {\bibfnamefont {E.}~\bibnamefont {Harms}},
  \bibinfo {author} {\bibfnamefont {M.}~\bibnamefont {Bugner}}, \bibinfo
  {author} {\bibfnamefont {H.}~\bibnamefont {R\"uter}}, \ and\ \bibinfo
  {author} {\bibfnamefont {B.}~\bibnamefont {Br\"ugmann}},\ }\href {\doibase
  10.1088/1361-6382/aaa4ac} {\bibfield  {journal} {\bibinfo  {journal} {Class.
  Quant. Grav.}\ }\textbf {\bibinfo {volume} {35}},\ \bibinfo {pages} {055003}
  (\bibinfo {year} {2018})},\ \Eprint {http://arxiv.org/abs/1609.08949}
  {arXiv:1609.08949 [gr-qc]} \BibitemShut {NoStop}%
\bibitem [{\citenamefont {Gautam}\ \emph {et~al.}(2021)\citenamefont {Gautam},
  \citenamefont {Va{\~n}{\'o}-Vi{\~n}uales}, \citenamefont {Hilditch},\ and\
  \citenamefont {Bose}}]{Gautam:2021ilg}%
  \BibitemOpen
  \bibfield  {author} {\bibinfo {author} {\bibfnamefont {S.}~\bibnamefont
  {Gautam}}, \bibinfo {author} {\bibfnamefont {A.}~\bibnamefont
  {Va{\~n}{\'o}-Vi{\~n}uales}}, \bibinfo {author} {\bibfnamefont
  {D.}~\bibnamefont {Hilditch}}, \ and\ \bibinfo {author} {\bibfnamefont
  {S.}~\bibnamefont {Bose}},\ }\href {\doibase 10.1103/PhysRevD.103.084045}
  {\bibfield  {journal} {\bibinfo  {journal} {Phys. Rev. D}\ }\textbf {\bibinfo
  {volume} {103}},\ \bibinfo {pages} {084045} (\bibinfo {year} {2021})},\
  \Eprint {http://arxiv.org/abs/2101.05038} {arXiv:2101.05038 [gr-qc]}
  \BibitemShut {NoStop}%
\bibitem [{\citenamefont {Peterson}\ \emph {et~al.}(2023)\citenamefont
  {Peterson}, \citenamefont {Gautam}, \citenamefont {Rainho}, \citenamefont
  {Va{\~n}{\'o}-Vi{\~n}uales},\ and\ \citenamefont
  {Hilditch}}]{Peterson:2023bha}%
  \BibitemOpen
  \bibfield  {author} {\bibinfo {author} {\bibfnamefont {C.}~\bibnamefont
  {Peterson}}, \bibinfo {author} {\bibfnamefont {S.}~\bibnamefont {Gautam}},
  \bibinfo {author} {\bibfnamefont {I.}~\bibnamefont {Rainho}}, \bibinfo
  {author} {\bibfnamefont {A.}~\bibnamefont {Va{\~n}{\'o}-Vi{\~n}uales}}, \
  and\ \bibinfo {author} {\bibfnamefont {D.}~\bibnamefont {Hilditch}},\ }\href
  {\doibase 10.1103/PhysRevD.108.024067} {\bibfield  {journal} {\bibinfo
  {journal} {Phys. Rev. D}\ }\textbf {\bibinfo {volume} {108}},\ \bibinfo
  {pages} {024067} (\bibinfo {year} {2023})},\ \Eprint
  {http://arxiv.org/abs/2303.16190} {arXiv:2303.16190 [gr-qc]} \BibitemShut
  {NoStop}%
\bibitem [{\citenamefont {{Rinne}}(2025)}]{Rinne:2025}%
  \BibitemOpen
  \bibfield  {author} {\bibinfo {author} {\bibfnamefont {O.}~\bibnamefont
  {{Rinne}}},\ }\href {\doibase 10.1088/1361-6544/ae153e} {\bibfield  {journal}
  {\bibinfo  {journal} {Nonlinearity}\ }\textbf {\bibinfo {volume} {38}},\
  \bibinfo {eid} {105026} (\bibinfo {year} {2025})},\ \Eprint
  {http://arxiv.org/abs/2507.00674} {arXiv:2507.00674 [cs.NA]} \BibitemShut
  {NoStop}%
\bibitem [{\citenamefont {Reddy}\ \emph {et~al.}(2026)\citenamefont {Reddy},
  \citenamefont {Gautam},\ and\ \citenamefont {Kumar}}]{Reddy:2026dfz}%
  \BibitemOpen
  \bibfield  {author} {\bibinfo {author} {\bibfnamefont {A.}~\bibnamefont
  {Reddy}}, \bibinfo {author} {\bibfnamefont {S.}~\bibnamefont {Gautam}}, \
  and\ \bibinfo {author} {\bibfnamefont {P.}~\bibnamefont {Kumar}},\
  }\href@noop {} {\  (\bibinfo {year} {2026})},\ \Eprint
  {http://arxiv.org/abs/2606.02051} {arXiv:2606.02051 [gr-qc]} \BibitemShut
  {NoStop}%
\bibitem [{\citenamefont {Gundlach}\ \emph {et~al.}(1994)\citenamefont
  {Gundlach}, \citenamefont {Price},\ and\ \citenamefont
  {Pullin}}]{Gundlach:1993tp}%
  \BibitemOpen
  \bibfield  {author} {\bibinfo {author} {\bibfnamefont {C.}~\bibnamefont
  {Gundlach}}, \bibinfo {author} {\bibfnamefont {R.~H.}\ \bibnamefont {Price}},
  \ and\ \bibinfo {author} {\bibfnamefont {J.}~\bibnamefont {Pullin}},\ }\href
  {\doibase 10.1103/PhysRevD.49.883} {\bibfield  {journal} {\bibinfo  {journal}
  {Phys.Rev.}\ }\textbf {\bibinfo {volume} {D49}},\ \bibinfo {pages} {883}
  (\bibinfo {year} {1994})},\ \Eprint {http://arxiv.org/abs/gr-qc/9307009}
  {arXiv:gr-qc/9307009 [gr-qc]} \BibitemShut {NoStop}%
\bibitem [{\citenamefont {Donninger}\ and\ \citenamefont
  {Zengino{\u{g}}lu}(2014)}]{Donninger:2013sba}%
  \BibitemOpen
  \bibfield  {author} {\bibinfo {author} {\bibfnamefont {R.}~\bibnamefont
  {Donninger}}\ and\ \bibinfo {author} {\bibfnamefont {A.}~\bibnamefont
  {Zengino{\u{g}}lu}},\ }\href {\doibase 10.2140/apde.2014.7.461} {\bibfield
  {journal} {\bibinfo  {journal} {Anal. Part. Diff. Eq.}\ }\textbf {\bibinfo
  {volume} {7}},\ \bibinfo {pages} {461} (\bibinfo {year} {2014})},\ \Eprint
  {http://arxiv.org/abs/1304.4135} {arXiv:1304.4135 [math.AP]} \BibitemShut
  {NoStop}%
\bibitem [{\citenamefont {{Burtscher}}\ and\ \citenamefont
  {{Donninger}}(2015)}]{Burtscher:2015}%
  \BibitemOpen
  \bibfield  {author} {\bibinfo {author} {\bibfnamefont {A.~Y.}\ \bibnamefont
  {{Burtscher}}}\ and\ \bibinfo {author} {\bibfnamefont {R.}~\bibnamefont
  {{Donninger}}},\ }\href {\doibase 10.48550/arXiv.1511.08600} {\bibfield
  {journal} {\bibinfo  {journal} {arXiv e-prints}\ ,\ \bibinfo {eid}
  {arXiv:1511.08600}} (\bibinfo {year} {2015})},\ \Eprint
  {http://arxiv.org/abs/1511.08600} {arXiv:1511.08600 [math.AP]} \BibitemShut
  {NoStop}%
\bibitem [{\citenamefont {{Bonk}}\ and\ \citenamefont
  {{Donninger}}(2026)}]{Bonk:2026}%
  \BibitemOpen
  \bibfield  {author} {\bibinfo {author} {\bibfnamefont {A.}~\bibnamefont
  {{Bonk}}}\ and\ \bibinfo {author} {\bibfnamefont {R.}~\bibnamefont
  {{Donninger}}},\ }\href {\doibase 10.48550/arXiv.2603.01924} {\bibfield
  {journal} {\bibinfo  {journal} {arXiv e-prints}\ ,\ \bibinfo {eid}
  {arXiv:2603.01924}} (\bibinfo {year} {2026})},\ \Eprint
  {http://arxiv.org/abs/2603.01924} {arXiv:2603.01924 [math.AP]} \BibitemShut
  {NoStop}%
\bibitem [{\citenamefont {Bizon}\ and\ \citenamefont
  {Zenginoglu}(2009)}]{Bizon:2008zd}%
  \BibitemOpen
  \bibfield  {author} {\bibinfo {author} {\bibfnamefont {P.}~\bibnamefont
  {Bizon}}\ and\ \bibinfo {author} {\bibfnamefont {A.}~\bibnamefont
  {Zenginoglu}},\ }\href@noop {} {\bibfield  {journal} {\bibinfo  {journal}
  {Nonlinearity}\ }\textbf {\bibinfo {volume} {22}},\ \bibinfo {pages} {2473}
  (\bibinfo {year} {2009})},\ \Eprint {http://arxiv.org/abs/0811.3966}
  {arXiv:0811.3966 [math.AP]} \BibitemShut {NoStop}%
\bibitem [{\citenamefont {Penrose}(1963)}]{Penrose:1962ij}%
  \BibitemOpen
  \bibfield  {author} {\bibinfo {author} {\bibfnamefont {R.}~\bibnamefont
  {Penrose}},\ }\href {\doibase 10.1103/PhysRevLett.10.66} {\bibfield
  {journal} {\bibinfo  {journal} {Phys. Rev. Lett.}\ }\textbf {\bibinfo
  {volume} {10}},\ \bibinfo {pages} {66} (\bibinfo {year} {1963})}\BibitemShut
  {NoStop}%
\bibitem [{\citenamefont {Frauendiener}(2000)}]{Frauendiener:2000mk}%
  \BibitemOpen
  \bibfield  {author} {\bibinfo {author} {\bibfnamefont {J.}~\bibnamefont
  {Frauendiener}},\ }\href@noop {} {\bibfield  {journal} {\bibinfo  {journal}
  {Living Rev.Rel.}\ }\textbf {\bibinfo {volume} {3}},\ \bibinfo {pages} {4}
  (\bibinfo {year} {2000})}\BibitemShut {NoStop}%
\bibitem [{\citenamefont {Gowdy}(1981)}]{Gowdy:1981}%
  \BibitemOpen
  \bibfield  {author} {\bibinfo {author} {\bibfnamefont {R.~H.}\ \bibnamefont
  {Gowdy}},\ }\href@noop {} {\bibfield  {journal} {\bibinfo  {journal} {Journal
  of Mathematical Physics}\ }\textbf {\bibinfo {volume} {22}},\ \bibinfo
  {pages} {675} (\bibinfo {year} {1981})}\BibitemShut {NoStop}%
\bibitem [{\citenamefont {Moncrief}(2000)}]{Moncrief:2000}%
  \BibitemOpen
  \bibfield  {author} {\bibinfo {author} {\bibfnamefont {V.}~\bibnamefont
  {Moncrief}},\ }\href@noop {} {\enquote {\bibinfo {title} {Conformally regular
  {ADM} evolution equations},}\ } (\bibinfo {year} {2000}),\ \bibinfo {note}
  {talk at Santa Barbara,
  \texttt{http://online.itp.ucsb.edu/online/numrel00/moncrief}}\BibitemShut
  {NoStop}%
\bibitem [{\citenamefont {Fodor}\ and\ \citenamefont
  {Racz}(2004)}]{Fodor:2003yg}%
  \BibitemOpen
  \bibfield  {author} {\bibinfo {author} {\bibfnamefont {G.}~\bibnamefont
  {Fodor}}\ and\ \bibinfo {author} {\bibfnamefont {I.}~\bibnamefont {Racz}},\
  }\href {\doibase 10.1103/PhysRevLett.92.151801} {\bibfield  {journal}
  {\bibinfo  {journal} {Phys. Rev. Lett.}\ }\textbf {\bibinfo {volume} {92}},\
  \bibinfo {pages} {151801} (\bibinfo {year} {2004})},\ \Eprint
  {http://arxiv.org/abs/hep-th/0311061} {arXiv:hep-th/0311061} \BibitemShut
  {NoStop}%
\bibitem [{\citenamefont {Friedrich}(1983)}]{Friedrich:1983}%
  \BibitemOpen
  \bibfield  {author} {\bibinfo {author} {\bibfnamefont {H.}~\bibnamefont
  {Friedrich}},\ }\href@noop {} {\bibfield  {journal} {\bibinfo  {journal}
  {Comm. Math. Phys.}\ }\textbf {\bibinfo {volume} {91}},\ \bibinfo {pages}
  {445} (\bibinfo {year} {1983})}\BibitemShut {NoStop}%
\bibitem [{\citenamefont {Panosso~Macedo}(2024)}]{PanossoMacedo:2023qzp}%
  \BibitemOpen
  \bibfield  {author} {\bibinfo {author} {\bibfnamefont {R.}~\bibnamefont
  {Panosso~Macedo}},\ }\href {\doibase 10.1098/rsta.2023.0046} {\bibfield
  {journal} {\bibinfo  {journal} {Phil. Trans. Roy. Soc. Lond. A}\ }\textbf
  {\bibinfo {volume} {382}},\ \bibinfo {pages} {20230046} (\bibinfo {year}
  {2024})},\ \Eprint {http://arxiv.org/abs/2307.15735} {arXiv:2307.15735
  [gr-qc]} \BibitemShut {NoStop}%
\bibitem [{\citenamefont {Va{\~n}{\'o}-Vi{\~n}uales}\ and\ \citenamefont
  {Valente}(2024)}]{Vano-Vinuales:2024tat}%
  \BibitemOpen
  \bibfield  {author} {\bibinfo {author} {\bibfnamefont {A.}~\bibnamefont
  {Va{\~n}{\'o}-Vi{\~n}uales}}\ and\ \bibinfo {author} {\bibfnamefont
  {T.}~\bibnamefont {Valente}},\ }\href {\doibase 10.1007/s10714-024-03323-8}
  {\bibfield  {journal} {\bibinfo  {journal} {Gen. Rel. Grav.}\ }\textbf
  {\bibinfo {volume} {56}},\ \bibinfo {pages} {135} (\bibinfo {year} {2024})},\
  \Eprint {http://arxiv.org/abs/2408.08952} {arXiv:2408.08952 [gr-qc]}
  \BibitemShut {NoStop}%
\bibitem [{\citenamefont {Zengino{\u{g}}lu}(2025)}]{Zenginoglu:2025sft}%
  \BibitemOpen
  \bibfield  {author} {\bibinfo {author} {\bibfnamefont {A.}~\bibnamefont
  {Zengino{\u{g}}lu}},\ }\href {\doibase 10.1007/s10714-025-03410-4} {\bibfield
   {journal} {\bibinfo  {journal} {Gen. Rel. Grav.}\ }\textbf {\bibinfo
  {volume} {57}},\ \bibinfo {pages} {75} (\bibinfo {year} {2025})},\ \Eprint
  {http://arxiv.org/abs/2502.08581} {arXiv:2502.08581 [gr-qc]} \BibitemShut
  {NoStop}%
\bibitem [{\citenamefont {N{\"u}tzi}(2025)}]{Nutzi:2025kqc}%
  \BibitemOpen
  \bibfield  {author} {\bibinfo {author} {\bibfnamefont {A.}~\bibnamefont
  {N{\"u}tzi}},\ }\href@noop {} {\  (\bibinfo {year} {2025})},\ \Eprint
  {http://arxiv.org/abs/2510.01964} {arXiv:2510.01964 [gr-qc]} \BibitemShut
  {NoStop}%
\bibitem [{\citenamefont {Parikh}(2002)}]{Parikh:2002qh}%
  \BibitemOpen
  \bibfield  {author} {\bibinfo {author} {\bibfnamefont {M.~K.}\ \bibnamefont
  {Parikh}},\ }\href {\doibase 10.1016/S0370-2693(02)02701-6} {\bibfield
  {journal} {\bibinfo  {journal} {Phys. Lett. B}\ }\textbf {\bibinfo {volume}
  {546}},\ \bibinfo {pages} {189} (\bibinfo {year} {2002})},\ \Eprint
  {http://arxiv.org/abs/hep-th/0204107} {arXiv:hep-th/0204107} \BibitemShut
  {NoStop}%
\bibitem [{\citenamefont {{Szpak}}(2009)}]{Szpak:2009}%
  \BibitemOpen
  \bibfield  {author} {\bibinfo {author} {\bibfnamefont {N.}~\bibnamefont
  {{Szpak}}},\ }\href {\doibase 10.48550/arXiv.0907.4287} {\bibfield  {journal}
  {\bibinfo  {journal} {arXiv e-prints}\ ,\ \bibinfo {eid} {arXiv:0907.4287}}
  (\bibinfo {year} {2009})},\ \Eprint {http://arxiv.org/abs/0907.4287}
  {arXiv:0907.4287 [math-ph]} \BibitemShut {NoStop}%
\bibitem [{\citenamefont {{Szpak}}(2010)}]{Szpak:2010}%
  \BibitemOpen
  \bibfield  {author} {\bibinfo {author} {\bibfnamefont {N.}~\bibnamefont
  {{Szpak}}},\ }\href {\doibase 10.1063/1.3470957} {\bibfield  {journal}
  {\bibinfo  {journal} {Journal of Mathematical Physics}\ }\textbf {\bibinfo
  {volume} {51}},\ \bibinfo {pages} {082901} (\bibinfo {year} {2010})},\
  \Eprint {http://arxiv.org/abs/0909.1264} {arXiv:0909.1264 [math-ph]}
  \BibitemShut {NoStop}%
\bibitem [{\citenamefont {Angelopoulos}\ \emph {et~al.}(2018)\citenamefont
  {Angelopoulos}, \citenamefont {Aretakis},\ and\ \citenamefont
  {Gajic}}]{Angelopoulos:2016wcv}%
  \BibitemOpen
  \bibfield  {author} {\bibinfo {author} {\bibfnamefont {Y.}~\bibnamefont
  {Angelopoulos}}, \bibinfo {author} {\bibfnamefont {S.}~\bibnamefont
  {Aretakis}}, \ and\ \bibinfo {author} {\bibfnamefont {D.}~\bibnamefont
  {Gajic}},\ }\href {\doibase 10.1016/j.aim.2017.10.027} {\bibfield  {journal}
  {\bibinfo  {journal} {Adv. Math.}\ }\textbf {\bibinfo {volume} {323}},\
  \bibinfo {pages} {529} (\bibinfo {year} {2018})},\ \Eprint
  {http://arxiv.org/abs/1612.01566} {arXiv:1612.01566 [math.AP]} \BibitemShut
  {NoStop}%
\bibitem [{\citenamefont {Luk}\ and\ \citenamefont {Oh}(2024)}]{Luk:2024ghv}%
  \BibitemOpen
  \bibfield  {author} {\bibinfo {author} {\bibfnamefont {J.}~\bibnamefont
  {Luk}}\ and\ \bibinfo {author} {\bibfnamefont {S.-J.}\ \bibnamefont {Oh}},\
  }\href@noop {} {\  (\bibinfo {year} {2024})},\ \Eprint
  {http://arxiv.org/abs/2404.02220} {arXiv:2404.02220 [gr-qc]} \BibitemShut
  {NoStop}%
\bibitem [{\citenamefont {N{\"u}tzi}(2023)}]{Nutzi:2023uly}%
  \BibitemOpen
  \bibfield  {author} {\bibinfo {author} {\bibfnamefont {A.}~\bibnamefont
  {N{\"u}tzi}},\ }\emph {\bibinfo {title} {{Maurer-Cartan perturbation theory
  and scattering amplitudes in general relativity}}},\ \href {\doibase
  10.3929/ethz-b-000625781} {Ph.D. thesis},\ \bibinfo  {school} {Zurich, ETH}
  (\bibinfo {year} {2023})\BibitemShut {NoStop}%
\bibitem [{\citenamefont {Zengino{\u g}lu}\ and\ \citenamefont
  {Tiglio}(2009)}]{Zenginoglu:2009hd}%
  \BibitemOpen
  \bibfield  {author} {\bibinfo {author} {\bibfnamefont {A.}~\bibnamefont
  {Zengino{\u g}lu}}\ and\ \bibinfo {author} {\bibfnamefont {M.}~\bibnamefont
  {Tiglio}},\ }\href {\doibase 10.1103/PhysRevD.80.024044} {\bibfield
  {journal} {\bibinfo  {journal} {Phys. Rev.}\ }\textbf {\bibinfo {volume}
  {D80}},\ \bibinfo {pages} {024044} (\bibinfo {year} {2009})},\ \Eprint
  {http://arxiv.org/abs/0906.3342} {arXiv:0906.3342 [gr-qc]} \BibitemShut
  {NoStop}%
\bibitem [{\citenamefont {{Burns}}\ \emph {et~al.}(2020)\citenamefont
  {{Burns}}, \citenamefont {{Vasil}}, \citenamefont {{Oishi}}, \citenamefont
  {{Lecoanet}},\ and\ \citenamefont {{Brown}}}]{Burns:2020}%
  \BibitemOpen
  \bibfield  {author} {\bibinfo {author} {\bibfnamefont {K.~J.}\ \bibnamefont
  {{Burns}}}, \bibinfo {author} {\bibfnamefont {G.~M.}\ \bibnamefont
  {{Vasil}}}, \bibinfo {author} {\bibfnamefont {J.~S.}\ \bibnamefont
  {{Oishi}}}, \bibinfo {author} {\bibfnamefont {D.}~\bibnamefont {{Lecoanet}}},
  \ and\ \bibinfo {author} {\bibfnamefont {B.~P.}\ \bibnamefont {{Brown}}},\
  }\href {\doibase 10.1103/PhysRevResearch.2.023068} {\bibfield  {journal}
  {\bibinfo  {journal} {Physical Review Research}\ }\textbf {\bibinfo {volume}
  {2}},\ \bibinfo {eid} {023068} (\bibinfo {year} {2020})},\ \Eprint
  {http://arxiv.org/abs/1905.10388} {arXiv:1905.10388 [astro-ph.IM]}
  \BibitemShut {NoStop}%
\bibitem [{\citenamefont {{Lecoanet}}\ \emph {et~al.}(2019)\citenamefont
  {{Lecoanet}}, \citenamefont {{Vasil}}, \citenamefont {{Burns}}, \citenamefont
  {{Brown}},\ and\ \citenamefont {{Oishi}}}]{Lecoanet:2019}%
  \BibitemOpen
  \bibfield  {author} {\bibinfo {author} {\bibfnamefont {D.}~\bibnamefont
  {{Lecoanet}}}, \bibinfo {author} {\bibfnamefont {G.~M.}\ \bibnamefont
  {{Vasil}}}, \bibinfo {author} {\bibfnamefont {K.~J.}\ \bibnamefont
  {{Burns}}}, \bibinfo {author} {\bibfnamefont {B.~P.}\ \bibnamefont
  {{Brown}}}, \ and\ \bibinfo {author} {\bibfnamefont {J.~S.}\ \bibnamefont
  {{Oishi}}},\ }\href {\doibase 10.1016/j.jcpx.2019.100012} {\bibfield
  {journal} {\bibinfo  {journal} {Journal of Computational Physics: X}\
  }\textbf {\bibinfo {volume} {3}},\ \bibinfo {pages} {100012} (\bibinfo {year}
  {2019})}\BibitemShut {NoStop}%
\bibitem [{\citenamefont {Vasil}\ \emph {et~al.}(2018)\citenamefont {Vasil},
  \citenamefont {Lecoanet}, \citenamefont {Burns}, \citenamefont {Oishi},\ and\
  \citenamefont {Brown}}]{Vasil:2018rsw}%
  \BibitemOpen
  \bibfield  {author} {\bibinfo {author} {\bibfnamefont {G.}~\bibnamefont
  {Vasil}}, \bibinfo {author} {\bibfnamefont {D.}~\bibnamefont {Lecoanet}},
  \bibinfo {author} {\bibfnamefont {K.}~\bibnamefont {Burns}}, \bibinfo
  {author} {\bibfnamefont {J.}~\bibnamefont {Oishi}}, \ and\ \bibinfo {author}
  {\bibfnamefont {B.}~\bibnamefont {Brown}},\ }\href@noop {} {\  (\bibinfo
  {year} {2018})},\ \Eprint {http://arxiv.org/abs/1804.10320} {arXiv:1804.10320
  [math.NA]} \BibitemShut {NoStop}%
\end{thebibliography}
\end{document}